\documentclass[12pt]{article}
\usepackage{amsmath,amssymb,amsfonts, amsbsy, amsthm, epsfig, subfig,natbib,soul,hyperref,enumerate}
\usepackage{color,rotating,multirow,algorithm,algpseudocode,booktabs}   

\newtheorem{theorem}{Theorem}
\newtheorem{corollary}{Corollary}
\newtheorem{proposition}{Proposition} 
\newtheorem{lemma}{Lemma}
\newtheorem{example}{Example} 

\newtheorem{definition}{Definition}

\newcommand{\beq}{\begin{equation}}
\newcommand{\eeq}{\end{equation}}
\newcommand{\beas}{\begin{eqnarray*}}
\newcommand{\eeas}{\end{eqnarray*}}
\newcommand{\bea}{\begin{eqnarray}}
\newcommand{\eea}{\end{eqnarray}}
\newcommand{\bei}{\begin{itemize}}
\newcommand{\eei}{\end{itemize}}
\newcommand{\ben}{\begin{enumerate}}
\newcommand{\een}{\end{enumerate}}
\newcommand{\bet}{\begin{theorem}}
\newcommand{\eet}{\end{theorem}}
\newcommand{\bel}{\begin{lemma}}
\newcommand{\eel}{\end{lemma}}
\newcommand{\bep}{\begin{proposition}}
\newcommand{\eep}{\end{proposition}}
\newcommand{\bed}{\begin{definition}}
\newcommand{\eed}{\end{definition}}
\newcommand{\bec}{\begin{corollary}}
\newcommand{\eec}{\end{corollary}}
\newcommand{\bex}{\begin{example}}
\newcommand{\eex}{\end{example}}

\def\A{\boldsymbol{A}}

\def\G{\boldsymbol{G}}

\def\R{\boldsymbol{R}}
\def\S{\boldsymbol{S}}

\def\X{\boldsymbol{X}}

\def\s{\boldsymbol{s}}

\def\bGamma{\boldsymbol{\Gamma}}
\def\bLambda{\boldsymbol{\Lambda}}

\def\bSigma{\boldsymbol{\Sigma}}
\def\0{\boldsymbol{0}}

\def\cF{\mathcal{F}}
\def\cG{\mathcal{G}}
\def\cH{\mathcal{H}}
\def\cI{\mathcal{I}}
\def\cJ{\mathcal{J}}
\def\cM{\mathcal{M}}

\def\cU{\mathcal{U}}

\def\pr{\textsf{P}} 
\def\ep{\textsf{E}} 
 
\def\Var{\textsf{Var}} 

\newcommand{\RR}{\mathbb{R}}

\def\T{{ \mathsf{\scriptscriptstyle T} }}

\begin{document}

\title{\Large{{\bf Hypothesis Testing for Network Data \\ with Power Enhancement}}}

\vspace{-0.2in}
\author{Yin Xia and Lexin Li \\
{\it Fudan University and University of California at Berkeley} }

\date{}

\maketitle

\begin{abstract}
Comparing two population means of network data is of paramount importance in a wide range of scientific applications. Many existing network inference solutions focus on global testing of entire networks, without comparing individual network links. Besides, the observed data often take the form of vectors or matrices, and the problem is formulated as comparing two covariance or precision matrices under a normal or matrix normal distribution. Moreover, many tests suffer from a limited power under a small sample size. In this article, we tackle the problem of network comparison, both global and simultaneous inferences, when the data come in a different format, i.e., in the form of a collection of symmetric matrices, each of which encodes the network structure of an individual subject. Such data format commonly arises in applications such as brain connectivity analysis and clinical genomics. We no longer require the underlying data to follow a normal distribution, but instead impose some moment conditions that are easily satisfied for numerous types of network data. Furthermore, we propose a power enhancement procedure, and show that it can control the false discovery, while it has the potential to substantially enhance the power of the test. We investigate the efficacy of our testing procedure through both an asymptotic analysis and a simulation study under a finite sample size. We further illustrate our method with an example of brain structural connectivity analysis.  
\end{abstract}


\newpage

\section{Introduction}
\label{intro.sec}

With prevalence of network data in recent years, the problem of comparing two populations of networks is gaining increasing attention. Our motivation is brain connectivity analysis, which studies functional and structural brain architectures through neurophysiological measures of brain activities and synchronizations \citep{Varoquaux2013}. Accumulated evidences have suggested that, compared to a healthy brain, the brain connectivity network alters in the presence of numerous neurological disorders, for example, Alzheimer's disease, autism spectrum disorder, among many others. Such alternations are believed to hold crucial insights of disease pathologies \citep{Fox2010}. A typical brain connectivity study collects imaging scans, such as functional magnetic resonance imaging, or diffusion tensor imaging, from groups of subjects with and without disorder. Based on the imaging scan, a network is constructed for each individual subject, with the nodes corresponding to a common set of brain regions, and the edges encoding the functional or structural associations between the regions. A fundamental scientific question of interest is to compare the brain networks and to identify local connectivity patterns that alter between the two populations. Network comparison is equally interesting in many other scientific areas as well, for instance, clinical genomics, where of crucial interest is to understand and compare gene regulatory networks of patients with and without cancer \citep{Luscombe2004}. 

In the context of brain connectivity analysis, there has been a rich literature on network \emph{estimation} methods \citep[among many others]{Ahn2015, QiuHan2016, Han2016, WangKang2016, ZhuLi2018}. There is, however, a relative paucity of \emph{inference} methods, especially simultaneous inference for individual links. Even though both can produce, in effect, a concise representation of the network structure, network inference is a fundamentally different problem than network estimation. Among the few existing network inference solutions, \citet{Kim2014} first summarized the network through a set of network metrics then employed a standard two-sample test. This strategy is commonly employed in the neuroscience literature and is easy to implement. However, it remains unclear to what extent each network metric provides a meaningful representation of brain function and structure \citep{Fornito2013}. \citet{ginestet2017hypothesis} characterized the geometry of the space of   labeled, undirected networks with edge weights, established a central limit theorem for an appropriate notion of a network empirical mean, then developed an analog of the classical two-sample test. \citet{kolaczyk2017averages} further extended this idea to unlabeled networks. 
However, these methods focused on the comparison of two entire networks and developed some global tests. None addressed the inference of individual links of two networks, nor considered simultaneous tests with false discovery control. \citet{ChenKang2015} developed a method to detect differentially expressed connectivity subnetworks under different clinical conditions by searching clusters of the graph. They resorted to a permutation test to obtain the $p$-value of the selected subnetwork, but did not provide the statistical significance of individual links and their differences. Besides, they only controlled the family-wise error rate, instead of the false discovery rate. \citet{xia2015testing} first encoded the connectivity network by a partial correlation matrix computed from vector-valued data under a normal distribution. They then proposed a multiple testing procedure to compare the partial correlation matrices from the two populations, along with a proper false discovery control. \citet{xia2018two} further extended the test to matrix-valued data under a matrix normal distribution. In both cases, the test statistics were constructed based on the vector or matrix-valued data, which, as we explain next, may not be directly observable. Moreover, the underlying data distribution may not always be normal or matrix normal. \citet{durante2018bayesian} developed a fully Bayesian solution for network comparison, which is very flexible and can handle the data format of our problem, but it requires specification of a series of prior distributions and can be computationally intensive. 

Applications such as brain connectivity analysis actually raise new challenges for network inference. First, the observed data come in the form of $p \times p$ matrices, each of which encodes the network structure for one individual subject, and $p$ is the number of network nodes. For instance, in brain structural connectivity, what one observes are the numbers of white matter fibers between pairs of brain anatomical regions, and this matrix of counts forms a network, with brain regions constituting the nodes and the fiber counts the links. This is ultimately different from the data format studied in most network methods, where a network structure is inferred from some vector-valued or matrix-valued data and usually takes the form a Pearson correlation or partial correlation matrix. This fundamental difference in terms of the available data format would thus require a completely new problem formulation and inferential procedure. Second, in a multitude of applications including brain connectivity analysis, the sample size is usually very small, e.g., in tens. This calls for a testing procedure that is powerful enough to detect differentially expressed links under a limited sample size. In this article, we address the problem of comparing two populations of network data, more precisely, the two population means of networks. We aim to consider both global and simultaneous inferences, tackle the new data format, and explicitly enhance the power of the test. 

Specifically, suppose we observe two groups of $p \times p$ network data, $\{\S_{1,1}, \ldots, \S_{1,n_1}\}$ and $\{\S_{2,1}, \ldots, \S_{2,n_2}\}$, where $n_1$, $n_2$ are the number of network samples for the two groups, respectively. Suppose $\S_{d,l} = (S_{d,l,i,j})_{p \times p} \sim \cF_d(\s_d)$, $l=1,\ldots,n_d, d=1,2$, where  $\cF_d$ is some distribution with a symmetric mean matrix $\s_d=(s_{d,i,j})_{p\times p}$. Our goal is to test whether the two population means are the same; i.e.,  
\begin{eqnarray}\label{global2}
H_0: \s_1=\s_2 \;\; \mbox{ versus } \;\; H_1: \s_1 \neq \s_2.
\end{eqnarray}
If the global null in \eqref{global2} is rejected, we further aim to identify at which locations the two mean matrices are different. That is, we wish to simultaneously test, 
\begin{eqnarray}\label{multiple2}
H_{0,i,j}: s_{1,i,j}=s_{2,i,j} \;\; \mbox{ versus } \;\; H_{1,i,j}: s_{1,i,j}\neq s_{2,i,j}, \; \mbox{ for } \; 1\leq i< j\leq p.
\end{eqnarray}
In \citet{xia2015testing}, the observed data, $\X_{d,l} \in \RR^{p}$, $l=1,\ldots,n_d$, $d=1,2$, is of the vector form, and is assumed to follow a normal distribution with the covariance matrix $\bSigma_d$. Let $\R_d$ denote the corresponding partial correlation matrix, i.e., the standardized version of $\bSigma_d^{-1}$, $d = 1, 2$. The network structure is  then encoded by $\R_d$, and the problem becomes testing if $\R_1 = \R_2$. \citet{xia2018two} followed a similar setup, except that the observed data $\X_{d,l} \in \RR^{p \times t}$ becomes a matrix, and is assumed to follow a matrix normal distribution with the covariance $\bSigma_d \otimes \bLambda_d$, and the network is still encoded by the standardized version of $\bSigma_d^{-1}$. The key difference for our setting is that, we do \emph{not} always observe $\X_{d,l}$ directly, but instead $\S_{d,l}$ \emph{only}. This difference in data format completely distinguishes our method from nearly all existing solutions such as \citet{xia2015testing} and \citet{xia2018two}. Moreover, we do not impose that the underlying data follows a normal or matrix normal distribution. Instead, we consider a general class of distributions for $\cF_d$ satisfying some moment condition. Our method works for many different types of network links, for instance, binary links when $\cF_d$ follows a light tailed distribution, or count links when $\cF_d$ follows a heavy-tailed distribution.   

For the global test \eqref{global2}, we develop a global test statistic taken as the maximum of a set of individual test statistics. We then derive its limiting null distribution, and show the resulting global test is power minimax optimal asymptotically. For the simultaneous test \eqref{multiple2}, we first develop a multiple testing procedure, and show that it can asymptotically control the false discovery at the pre-specified level. Next we propose a method to substantially enhance the power of the simultaneous inference procedure for \eqref{multiple2}. Specifically, we extend the grouping-adjusting-pooling idea of \citet{Xia2018GAP}, and modify it for our inference of network data. 

Our proposal differs from the existing solutions and makes several useful contributions. First, to the best of our knowledge, there has been no solution directly targeting simultaneous hypothesis testing of individual links for the network data in the format of $\S_{d,l}$. Our method bridges this gap, and offers a timely solution to a range of scientific applications where this form of problem and data is commonly encountered. Second, 
our global test statistic is constructed as the maximum of the individual test statistics for all links. This type of maximum statistic enjoys various advantages and has been commonly employed in the hypothesis testing literature \citep[e.g.,][]{cai2013two, cai2016inference, xia2019simultaneous}. However, the derivation of its asymptotics, as well as the properties of the subsequent multiple testing procedure, are far from trivial in our new context of network comparison. Moreover, we remark that, in some network data applications, the individual test statistics may be correlated, and a global test statistic that utilizes such correlations may result in a more powerful test. However, this may not always be the case. For instance, in our brain connectivity application, the nodes are usually the brain anatomical regions, which can scatter at distant locations of the brain. As a result, there is no obvious correlation structure for the individual test statistics built on the pairs of brain regions. Therefore, we do not explicitly impose or employ any correlation structure when constructing the global test statistic. On the other hand, in our power enhancement procedure, we implicitly utilize the fact that some individual test statistics may be correlated and clustered. We then use a data driven approach to find such clusters and incorporate this  information in our test. Finally, the power enhancement approach we develop is particularly useful in many applications, e.g., brain connectivity analysis, where the sample size is limited. Although motivated by \citet{Xia2018GAP}, our enhancement method differs from \citet{Xia2018GAP} considerably in several ways. We explicitly compare the two power enhancement procedures in Section \ref{compare.sec}. Overall, we feel our method provides a useful addition to the general toolbox of network inference. 

We adopt the following notation throughout this article. For a symmetric matrix $\A_d$,  let $\lambda_{\max}(\A_d)$ and $\lambda_{\min}(\A_d)$ denote the largest and smallest eigenvalues of $\A_d$, respectively. For a set $\mathcal{H}$, let $|\mathcal{H}|$ denote its cardinality. For two  sequences of real numbers $\{a_{n}\}$ and $\{b_{n}\}$, write $a_{n} = O(b_{n})$ if there exists a constant $C$ such that $|a_{n}| \leq C|b_{n}|$ holds for all $n$, write $a_{n} = o(b_{n})$ if $\lim_{n\rightarrow\infty}a_{n}/b_{n} = 0$, and write $a_{n}\asymp b_{n}$ if there are positive constants $c$ and $C$ such that $c\leq a_{n}/b_{n}\leq C$ for all $n$.  Write $n = {n_1 n_2} / {(n_1 + n_2)}$ and assume that $n_1\asymp n_2$.

The rest of the article is organized as follows. Section \ref{assumption.sec} presents the moment conditions for the distribution of $\cF_d$ and show they are easily satisfied in numerous types of network data. Section \ref{method.sec} develops the global testing and the simultaneous testing procedures for the two-sample network comparison, and Section \ref{power.sec} studies power enhancement, both of which are key to our proposal. Section \ref{numerical.sec} presents the simulations, and a brain connectivity analysis as an illustration. The Supplementary Material collects additional lemmas and the proofs.

\section{Moment Conditions and Examples}
\label{assumption.sec}

We begin with some moment conditions imposed on $\cF_d$. We then give a number of examples and show that those conditions are easily satisfied in numerous types of network data.

\subsection{Moment conditions}
\label{moment.sec}

We assume that the distribution $\cF_d$ of the network data $\S_{d,l}$ satisfies one of the following two conditions: a sub-Gaussian-type tail, or a polynomial-type tail, as stated below. 

\begin{enumerate}[(C1)]
\item (Sub-Gaussian-tail). Suppose that $\log p=o(n^{1/5})$, and that there exist some constants $\eta>0$ and $K>0$, such that, for $d=1,2$,
\begin{eqnarray*}
\ep\left[ \exp\left\{ \eta (S_{d,l,i,j} - s_{d,i,j})^2/\Var (S_{d,l,i,j}) \right\} \right] \leq K, \;\; \textrm{ for } \; 1\leq i<j\leq p, \; l=1, \ldots, n_d. 
\end{eqnarray*}

\item (Polynomial-tail). Suppose that $p\leq cn^{\gamma_0}$ for some constants $\gamma_0$ and $c>0$, and that there exist some constants $\epsilon>0$ and $K>0$, such that, for $d=1,2$,
\begin{eqnarray*}
\ep\left\{ \left| (S_{d,l,i,j} - s_{d,i,j})/\Var (S_{d,l,i,j})^{1/2} \right|^{{ 4}\gamma_0+2+\epsilon} \right\} \leq K, \;\; \textrm{ for } \; 1\leq i<j\leq p, \; l=1, \ldots, n_d.
\end{eqnarray*}
\end{enumerate} 

\noindent
We first comment that, both conditions are common, and similar conditions have been often assumed in the high-dimensional setting \citep{cai2014two,van2014asymptotically}. These moment conditions are much weaker than the Gaussian assumption as usually required in the testing literature \citep{schott2007some, srivastava2010testing}. Next we discuss a number of network data examples that satisfy the above moment conditions, including Bernoulli and mixture Bernoulli data, Poisson data, as well as correlation and partial correlation data. Furthermore, we discuss some examples where the distributions are heavy-tailed, but after some data transformation, they still satisfy the moment conditions. Examples include transformed normal count data and transformed Wishart count data.

\subsection{Binary network data}\label{binary.sec}

Binary network is arguably the most commonly seen network data type, where each link is a binary indicator. The Bernoulli distribution is often assumed; i.e., for $\S_{d,l} = (S_{d,l,i,j})_{p \times p}$, $S_{d,l,i,j}$ follows a Bernoulli distribution with mean $s_{d,i,j}$, $u<s_{d,i,j}<1-u$ for a constant $0<u<1$, $l=1,\ldots,n_d, d=1,2$, and $1\leq i<j\leq p$. In such case, $\S_{d,l}$ satisfies the sub-Gaussian-tail condition in (C1), e.g., with $\eta=1$ and $K=(1-u)\exp\{ u(1-u)^{-1} \} + u\exp\{ (1-u)u^{-1} \}$. 

The same holds true for the mixture Bernoulli distribution as discussed in \cite{durante2018bayesian}. That is, for some integer $H>0$ and randomly selected $\{\phi_1,\ldots,\phi_H\}$ subject to $\sum_{h=1}^H\phi_h=1$ and $\phi_h>0$, $\pr(S_{d,l,i,j}=x)= \sum_{h=1}^H \phi_h \left\{ s^{(h)}_{d,i,j} \right\}^x \left\{ 1-s^{(h)}_{d,i,j} \right\}^{1-x}$, with $u<s^{(h)}_{d,i,j}<1-u$ for some constant $0<u<1$, $x=0,1$, $h=1,\ldots,H$, $l=1,\ldots,n_d, d=1,2$ and $1\leq i<j\leq p$. For this example, $\S_{d,l}$ again satisfies the sub-Gaussian-tail condition in (C1), with $\eta=1$ and $K=(1-u)\exp\{ u(1-u)^{-1} \} + u\exp\{ (1-u)u^{-1} \}$.

\subsection{Correlation network data}\label{correlation.sec}

Correlation network is another equally common network data type. In brain functional connectivity analysis and many other applications, the network is often encoded by a Pearson correlation or a partial correlation matrix. Take the Pearson correlation network as an example. The functional imaging data is usually summarized as a spatial-temporal matrix. That is, for the $l$th subject in the $d$th group, the observed data is of the form $\X_{d,l} \in \RR^{p \times t_d}$, $l=1,\ldots,n_d, d=1,2$, where $p$ is the number of brain regions, and $t_d$ is the number of repeated measures. Then the brain functional connectivity network is encoded by the sample correlation matrix $\S_{d,l} = t_d^{-1} \sum_{j=1}^{t_d} \{ \X_{d,l,(\cdot,j)} - \bar \X_{d,l} \} \{ \X_{d,l,(\cdot,j)} - \bar \X_{d,l} \}^{\T}$, where $\X_{d,l,(\cdot,j)}$ denotes the $j$th column of the matrix $\X_{d,l}$ and $\bar \X_{d,l} = t_d^{-1} \sum_{j=1}^{t_d}\X_{d,l,(\cdot,j)}$ denotes the sample mean vector \citep{Fornito2013}. Next we show that, as long as $\X_{d,l}$ satisfies one of the conditions in Lemma \ref{thm:correlation}, then $\S_{d,l}$ satisfies the sub-Gaussian-tail condition (C1).  

\begin{lemma} \label{thm:correlation}
Suppose $\X_{d,l}$ satisfies one of the following conditions: (i) $\log p=o(t^{1/5})$, and there exist constants $\eta' > 0, K' > 0$ such that $\ep\left( \exp\left[ \eta' \{X_{d,l,i,j} - \ep(X_{d,l,i,j})\}^{2}/\Var(X_{d,l,i,j}) \right] \right)$ $\leq K'$, where $t=\max\{t_1,t_2\}$ and $t_1\asymp t_2$; (ii) $p \leq c't^{\gamma_{0}'}$, for some $\gamma_{0}', c'>0$, and there exist constants $\epsilon' >0, K' > 0$ such that $\ep\left[ \left| \left\{ X_{d,l,i,j} - \ep(X_{d,l,i,j}) \right\} /\Var(X_{d,l,i,j})^{1/2} \right|^{4\gamma_{0}'+4+\epsilon'} \right] \leq K'$, for $i = 1,\ldots,p, j=1,\ldots,t_d$. Then $\S_{d,l}$ satisfies the sub-Gaussian-tail condition in (C1), with $\eta=1/4$ and $K=2$, as $t\rightarrow \infty$. 
\end{lemma}

\noindent
We remark that a similar result as Lemma \ref{thm:correlation} can be obtained for the partial correlation network, by using the inverse regression techniques as in \cite{liu2013ggm}. \citet{xia2018two} tackled the network comparison problem assuming $\X_{d,l}$ is directly observable and follows a matrix normal distribution. Lemma \ref{thm:correlation} suggests that, the test we develop later is still applicable when $\X_{d,l}$ is available, even though it may not be as powerful as the test of \citet{xia2018two} in this case. On the other hand, the main focus of this article is to develop a test of comparing two networks even when $\X_{d,l}$ is not observed, but only $\S_{d,l}$ is. As such, our test is more general than that of \citet{xia2018two}.

\DeclareRobustCommand{\stirling}{\genfrac\{\}{0pt}{}}
\subsection{Poisson network data}\label{poisson.sec}

Count network is another common network data type, where each link is a count. For instance, in brain structural connectivity analysis, the link is the number of white matter fibers between anatomical brain regions. The Poisson distribution is often imposed; i.e., $S_{d,l,i,j}$ follows a Poisson distribution with mean $s_{d,i,j}$, $0 < u_1<s_{d,i,j}<u_2$, $l=1,\ldots,n_d, d=1,2$, $1\leq i<j\leq p$. For any constant $\epsilon>0$, let $M$ be the smallest integer that is no smaller than $4\gamma_0+2+\epsilon$, where $\gamma_0$ is as defined in (C2). Then $\S_{d,l}$ satisfies the polynomial-tail condition (C2), with $K$ upper bounded by $u_1^{-(M-1)/2}\left[ \sum_{i=0}^Mu_2^i\stirling{M}{i}+u_2^M(u_2/2+1) \right]$, and $\stirling{M}{i}$ is the number of ways to partition a set of $M$ objects into $i$ non-empty subsets.

\subsection{Transformed network data}\label{log-normal.sec}

We next consider some examples where the original network data $\G_{d,l} = (G_{d,l,i,j})_{p \times p} \sim \tilde\cF_d(\tilde\s_d)$, $ l=1,\ldots,n_d, d=1,2$, and $\tilde\cF_d$ is some heavy-tailed distribution that only differs in the mean matrix $\tilde\s_d=(\tilde s_{d,i,j})\in \RR^{p\times p}$ between the two groups. In such cases, testing the means of the original samples are equivalent to testing the means of the transformed data, $S_{d,l,i,j}=f(G_{d,l,i,j})$, where $f$ is some one-to-one transformation function. We next give two examples, where after  transformation, the transformed data satisfies (C1) or (C2) or both.

One example is the log-normal count network. After the logarithmic transformation of $\G_{d,l}$, the transformed data $\S_{d,l}$ follows a normal distribution, and thus both (C1) and (C2) are satisfied. This can be further extended to the transformed normal mixture network.

Another example is the transformed Wishart count network, where the transformed data $\S_{d,l}$ follows the Wishart distribution with the scale matrix $m^{-1}\bSigma_{d}$ and the degrees of freedom $m$. For this case, $\S_{d,l}$ satisfies the sub-Gaussian-tail condition (C1). Moreover, in this case, the testing problems \eqref{global2} and \eqref{multiple2} are closely related to the covariance matrix testing problems studied in \cite{li2012two} and \cite{cai2013two}. The key difference between our method and the existing ones is that, we only observe $\S_{d, l}$, but not the original vector samples. This example can also be further extended to the case of the product of Gaussian mixtures network, or the Wishart mixtures network.

\section{Two-sample Test on Network Data}
\label{method.sec}

We begin with the construction of test statistic for the two testing problems \eqref{global2} and \eqref{multiple2}. We then develop a global testing procedure for \eqref{global2}, and a simultaneous testing procedure for \eqref{multiple2}. For each test, we derive its corresponding asymptotic properties.

\subsection{Test statistics}\label{stat.sec}

We first observe that the testing problem \eqref{global2} is equivalent to the test, $H_0': \max_{1\leq i<j\leq p}|s_{1,i,j}-s_{2,i,j}| = 0$. This motivates us to construct the test statistic based on
\begin{eqnarray*}
W_{i,j} = \bar S_{1,i,j} - \bar S_{2,i,j}.
\end{eqnarray*}
where $\bar S_{d,i,j} = n_d^{-1}\sum_{l=1}^{n_d}S_{d,l,i,j}$. We standardize $W_{i,j}$, and estimate the variance of $S_{d, l, i, j}$ by 
\begin{eqnarray}\label{Vdij}
V_{1,i,j} = n_1^{-1}\sum_{l=1}^{n_1}(S_{1,l,i,j} - \bar S_{1,i,j})^2, \; \textrm{ and } \; V_{2,i,j} = n_2^{-1}\sum_{l=1}^{n_2}(S_{2,l,i,j} - \bar S_{2,i,j})^2, 
\end{eqnarray}
respectively. This leads to our test statistic, 
\begin{eqnarray}\label{Tij}
T_{i,j} = \frac{W_{i,j}}{(V_{1,i,j}/n_1+V_{2,i,j}/n_2)^{1/2}},  \quad1\leq i< j\leq p.
\end{eqnarray}

\subsection{Global test}\label{global.sec}

In brain connectivity analysis and many other applications, it is generally postulated that the differences between two network structures concentrate on a small number of brain regions. This translates to a sparse alternative in our global test. Correspondingly, we construct the global test statistic as, 
\vspace{-0.1in}
\begin{eqnarray*}\label{global.test}
M_n = \max_{1\leq i<j\leq p} T^2_{i,j}.
\end{eqnarray*} 

Let $\bGamma_d \in \RR^{q\times q}$ denote the covariance matrix of vech($\S_{d,l}$), where $q=p(p-1)/2$, and vech$(\cdot)$ is the  operator that turns the upper triangular part of $\S_{d,l}$ into a vector. Let $\R_d=(r_{d,i,j})\in \RR^{q\times q}$ denote the corresponding correlation matrix. We introduce two conditions.
\begin{enumerate}[({A}1)]
\item $C_0^{-1}\leq \lambda_{\min}(\bGamma_d)\leq \lambda_{\max}(\bGamma_d)\leq C_0$ for some constant $C_0>0$, $d=1,2$.

\item $\max_{d=1,2}\max_{1\leq i<j\leq q}|r_{d,i,j}|< r<1$ for some constant $0<r<1$.
\end{enumerate}
Both conditions are mild. Particularly, Condition (A1) on the eigenvalues of the covariance matrix is common in the high dimensional setting \citep{bickel2008regularized, rothman2008sparse, yuan2010high, cai2014two}. Condition (A2) is also mild, because if $\max_{1\leq i<j\leq q}|r_{d,i,j}|=1$, then $\bGamma_d$ is singular. We next obtain the limiting distribution of our test statistic $M_n$.

\begin{theorem} \label{limit.null}
Suppose that (A1)-(A2), and one of (C1) and (C2) hold. Then for any $x\in \RR$,
\begin{eqnarray*}
\pr_{H_0}\left( M_n-2 \log q + \log\log q\leq x \right) \rightarrow \exp\left\{ -\pi^{-1/2}\exp(-x/2) \right\}, \;\; \textrm{as} \;\; n_1, n_2, q\rightarrow \infty. 
\end{eqnarray*}
\end{theorem}
Based on this limiting null distribution, we define the asymptotic $\alpha$-level test as, 
\begin{eqnarray*}\label{Psi}
\Psi_{\alpha} = I(M_n \geq 2 \log q - \log\log q + q_{\alpha}),
\end{eqnarray*}
where $q_{\alpha} =-\log \pi - 2\log \log (1-\alpha)^{-1}$.

We next study the power and the asymptotic optimality of the test $\Psi_{\alpha}$. Toward that end, define the sparsity of $\s_1-\s_2$ as $k_q = |\{(i,j): s_{1,i,j}-s_{2,i,j}\neq 0, 1\leq i<j\leq p\}|$. We also introduce a class of $(\s_1,\s_2)$, 
\begin{eqnarray*}
\cU(c) = \left\{ (\s_1,\s_2):~ \max_{1\leq i<j\leq p} \frac{ |s_{1,i,j}-s_{2,i,j}| }{ \left\{ \Var (S_{1,l,i,j})/n_1+\Var (S_{2,l,i,j})/n_2 \right\}^{1/2} }\geq c(\log q)^{1/2} \right\}.
\end{eqnarray*}

\begin{theorem} \label{power}
Suppose that one of (C1) and (C2) holds. Then, 
\begin{eqnarray*}
\inf_{(s_1,s_2)\in \cU(2\sqrt{2})} \pr\left( \Psi_{\alpha}=1 \right) \rightarrow 1, \;\; \textrm{as} \;\; n_1, n_2, q\rightarrow \infty. 
\end{eqnarray*}
Furthermore, suppose that $k_q = o(q^r)$ for some $r<1/2$. Let $\alpha,\beta>0$ and $\alpha+\beta=1$. Then there exists a constant $c_0>0$ such that, for all sufficiently large $n_d$ and $q$,
\begin{eqnarray*}
\inf_{(s_1,s_2)\in U(c_0)}\sup_{T_{\alpha}\in \mathcal{T}_{\alpha}} \pr\left( T_{\alpha}=1 \right) \leq 1-\beta,
\end{eqnarray*}
where $\mathcal{T}_{\alpha}$ is the set of all $\alpha$-level tests, i.e., $\pr_{H_0}({T}_{\alpha}=1)\leq\alpha$ for all $T_{\alpha}\in \mathcal{T}_{\alpha}$.
\end{theorem}
This theorem shows that the null hypothesis in \eqref{global2} can be rejected by $\Psi_{\alpha}$ with a high probability if the pair of the network means belong to the class $\cU(2\sqrt{2})$. In addition, with the mild sparsity condition $k_q = o(q^r)$, the lower bound rate of $(\log q)^{1/2}$ cannot be further improved, because for a sufficiently small $c_0$, any $\alpha$-level test is unable to reject the null correctly uniformly over $\cU(c_0)$ with probability tending to 1. Henceforth, the global test $\Psi_{\alpha}$ reaches the power minimax optimality asymptotically.

\subsection{Simultaneous test}\label{multiple.sec}

We next develop a multiple testing procedure for \eqref{multiple2} based on the test statistic $T_{i,j}$ in \eqref{Tij}. Let $h$ be the threshold level such that $H_{0,i,j}$ is rejected if $|T_{i,j}|\geq h$. Let $\cH_0=\{(i,j): s_{1,i,j}=s_{2,i,j},1\leq i<j\leq p\}$ be the set of true nulls, and $\cH_1=\cH\setminus \cH_0$ the set of true alternatives, where $\cH=\{(i,j):1\leq i<j\leq p\}$. Denote by $R_{0}(h) = \sum_{(i,j)\in \cH_0}I(|T_{i,j}|\geq h)$ and $R(h)= \sum_{1\leq i<j\leq p}I(|T_{i,j}|\geq h)$ the total number of false positives and rejections, respectively. Then we define the false discovery proportion and false discovery rate by
\begin{eqnarray*}
\textrm{FDP}(h)=\frac{R_{0}(h)}{R(h)\vee 1}, \quad  \textrm{FDR}(h)=\ep\{\textrm{FDP}(h)\}.
\end{eqnarray*}
An ideal choice of $h$ would reject as many true positives as possible while controlling the FDP at the pre-specified level $\alpha$. That is, we select $h_0=\inf\left\{h \, : \, 0\leq h\leq (2\log q)^{1/2}, \, \text{FDP}(h)\leq \alpha\right\}$. Since $R_{0}(h)$ is unknown, we estimate it conservatively by $2q\{1-\Phi(h)\}$, where $\Phi(h)$ is the standard normal cumulative distribution function. This leads to our multiple testing procedure as summarized in Algorithm \ref{proc1}.

\begin{algorithm}[t!]
\caption{Simultaneous inference with FDR control}
\label{proc1}
\begin{algorithmic}
\item Step 1: Estimate FDP by $\widehat{\text{FDP}}(h)=2q\{1-\Phi(h)\}/\{R(h)\vee 1\}$.
\item Step 2: For a given $0\leq \alpha\leq 1$, calculate
\[
\hat{h}=\inf\left\{h \, : \, 0\leq h\leq (2\log q)^{1/ 2}, \, \widehat{\text{FDP}}(h) \leq \alpha\right\}.
\]
\hspace{0.5in} If $\hat{h}$ does not exist, set $\hat{h}=(2\log q)^{1/ 2}$.
\item Step 3: Reject $H_{0,i,j}$ if and only if $|T_{i,j}|\geq \hat{h}$, for $1\leq i<j\leq p$. 
\end{algorithmic}
\end{algorithm}

We next show that this testing procedure controls the FDR and FDP asymptotically at the pre-specified level. For notation simplicity, we write FDP=FDP($\hat h$) and FDR=FDR($\hat h$), where $\hat h$ is obtained in Algorithm \ref{proc1}. Define $\mathcal{A}_i(\xi) = \{j: \max( |r_{1,i,j}|, |r_{2,i,j}|) \geq (\log q)^{-2-\xi}\}$, and $\mathcal{S}_{\rho} = \{(i,j) : 1\leq i<j\leq p, |s_{1,i,j}-s_{2,i,j}| / \{ \Var (S_{1,l,i,j})/n_1+\Var (S_{2,l,i,j})/n_2 \}^{1/2} \geq (\log q)^{1/2+\rho}\}$. We further introduce some conditions. 
\begin{enumerate}[({B}1)]
\item $|\mathcal{S}_{\rho} |\geq [{1}/\{\pi^{1/2}\alpha\}+\delta]({\log q})^{1/2}$, for some constant $\delta>0$ and any sufficiently small constant $\rho>0$. 

\item $\max_{1\leq i\leq q}|\mathcal{A}_i(\xi)| = o(q^{\nu})$ for some constants $\xi>0$ and $0<\nu<(1-r)/(1+r)$. 

\item $q_0= |\cH_0|\geq c_1 q$ for some constant $c_1>0$. 
\end{enumerate}
Condition (B1) on $\mathcal{S}_{\rho}$ is mild, as it only requires a small number of $\s_1$ and $\s_2$ having standardized difference with the order of $(\log q)^{1/2+\rho}$ for any sufficiently small constant $\rho>0$. Condition (B2) is mild, as it requires that not too many $S_{d,l,i,j}$ are highly correlated, but still allows the number of highly correlated pairs to grow in the order of $o(q^{1+\nu})$. Condition (B3) is also a natural and mild assumption, because if it does not hold, i.e., $q_0=o(q)$, then we can simply reject all the hypotheses. As a result, we would have $|R_0|=q_0$, $|R|=q$, and the FDR would tend to zero. Under these conditions, we obtain the asymptotic properties of our multiple testing procedure in terms of false discovery control. 

\begin{theorem} \label{FDR}
Suppose that (A2), (B1)-(B3), and one of (C1) and (C2) hold. Then, 
\begin{eqnarray*}
\lim_{(n_1,n_2,q)\rightarrow \infty}\frac{\textrm{FDR}}{\alpha q_0/q}=1,
\quad \textrm{and} \quad
\frac{\textrm{FDP}}{\alpha q_0/q} \rightarrow 1 \; \textrm{ in probability}, \;\; \textrm{as} \;\; n_1,n_2,q \rightarrow \infty.
\end{eqnarray*}
\end{theorem}

\section{Power Enhancement}
\label{power.sec}

In brain connectivity analysis and many other applications, the sample size $n_d$ is often small, whereas the number of nodes $p$ can be moderate to large. This results in a limited power for the proposed testing procedure. We explore in this section an explicit power enhancement method that has potential to substantially improve the power of the simultaneous inference developed in Section \ref{multiple.sec}. We borrow the idea of grouping, adjusting and pooling (GAP) that was first proposed in \citet{Xia2018GAP}. However, our method differs from \citet{Xia2018GAP} in many ways, including a different, and actually less restrictive, assumption, a different set of primary and auxiliary statistics, and a different modification of the multiple testing procedure. We show that the modified procedure is asymptotically more powerful, meanwhile it can still control FDR and FDP asymptotically. We obtain these properties assuming the sub-Gaussian-tail condition (C1). Parallel results can be obtained under the polynomial-tail condition (C2) as well, but are technically more involved. We begin by describing the intuition behind our power enhancement solution, then derive the proper auxiliary statistic for our inference problem. We then develop the modified simultaneous testing procedure, and study its asymptotic properties in terms of power improvement and false discovery control. We also compare in detail our method with the GAP method of \citet{Xia2018GAP}.

\subsection{Intuition}\label{intuition.sec}

We recognize that there exists additional information in the data that is potentially useful to improve the simultaneous testing procedure of Algorithm \ref{proc1}. We first discuss our intuition, then use some simple example to illustrate where the auxiliary information is and how it can facilitate our multiple testing procedure. 

In a multitude of applications including brain connectivity analysis, it is often believed that the difference between the two networks under different biological conditions is small. This means $\s_1 - \s_2$ is sparse. Accordingly, one can find a baseline matrix $\s_0$, such that $\s_1' = \s_1 - \s_0$ and $\s_2' = \s_2 - \s_0$ are individually sparse. Let $\mathcal I_d=\{(i,j): s_{d,i,j}'\neq 0, 1\leq i<j\leq p\}$ denote the support of $\s_d'$, $d=1, 2$, and $\cI=\cI_1 \cup \cI_2$ denote the union support. Note that the set of alternative hypotheses $\cH_1$ defined in Section \ref{multiple.sec} is the same as $\cI$, if $s_{1,i,j}\neq s_{2,i,j}$ for every $(i,j)\in\cI_1\cap\cI_2$. In general, $\cH_1$ is a proper subset of $\cI$. Since $\s_1'$ and $\s_2'$ are both sparse, we realize that the cardinality of $\cI$ is small. Moreover, the following relationship holds true: 
\begin{eqnarray*}\label{logical-relation}
(i,j)\notin \cI \;\; \textrm{ implies that } \; s_{1,i,j}-s_{2,i,j}=0, \quad 1\leq i<j\leq p. 
\end{eqnarray*}
Therefore, the knowledge about $\cI$ is useful to help narrow down the search in multiple testing. In other words, if one can find a way to identify possible entries $(i,j)$ in $\cI$, it would provide useful information about the set of true alternatives $\cH_1$, or equivalently, the set of true nulls $\cH_0$. As a consequence, it can potentially increase the power of the testing procedure. 

A key observation is then, while the test statistic is built on the difference between $\bar S_{1,i,j}$ and $\bar S_{2,i,j}$ as defined in Section \ref{stat.sec}, the sum of $\bar S_{1,i,j}$ and $\bar S_{2,i,j}$ can provide crucial information about $\mathcal I$. Consider a toy example where the network data is binary, and $S_{d,l,i,j}$ follows a Bernoulli distribution with mean $s_{d,l,i,j}$, $l=1, \ldots, n_d, d=1,2, 1\leq i<j\leq p$. Assume that $s_{1,i,j} = s_{2,i,j} = s_{0,i,j} = 0.1$ for $80\%$ of the $(i,j)$ pairs, $s_{1,i,j} = s_{2,i,j} = s_{0,i,j} = 0.9$ for $10\%$ of the $(i,j)$ pairs, and for the rest of the $(i,j)$ pairs, $s_{1,i,j},s_{2,i,j}\sim \text{Uniform}(0.1,0.9)$ and $s_{0,i,j}=0.1$.
In this example, for the pairs $(i,j) \notin \cI$, the sum of $s_{1,i,j}$ and $s_{2,i,j}$ is either very small, which is 0.2, or very large, i.e., 1.8. Meanwhile, for the pairs $(i,j) \in \cI$, the sum is in between. Henceforth, this sum contains useful information about $\cI$, and can potentially enhance the power of the multiple testing procedure.  

Based on the above discussion, we can see that, the more sparsity structure information the auxiliary statistics can capture, the more information they can provide about the union support $\cI$, and the more substantial power gain the test can achieve. In general, the sparser the true difference $\s_1-\s_2$ is, the more information the auxiliary statistics can offer.

\subsection{Auxiliary statistics}\label{auxiliary.sec}

We next formally construct the auxiliary statistic that provides useful information about the union support $\cI$. It is important to note that, the auxiliary statistic should be constructed so that they are asymptotically independent of the test statistic $T_{i,j}$ in \eqref{Tij}. This way the null distribution of $T_{i,j}$ would not be distorted by the incorporation of the auxiliary statistic.

Recall that $V_{d,i,j}$ is the sample variance of $S_{d,l,i,j}$ as defined in \eqref{Vdij}. We construct the auxiliary statistic $A_{i,j}$ as,
\begin{eqnarray*}
A_{i,j} = \frac{\bar S_{1,i,j}+\hat \kappa_{i,j}\bar S_{2,i,j}}{(V_{1,i,j}/n_1+\hat \kappa_{i,j}^2V_{2,i,j}/n_2)^{1/2}}, \quad1\leq i< j\leq p, 
\end{eqnarray*}
where $\hat \kappa_{i,j} = (n_2V_{1,i,j})/(n_1V_{2,i,j})$. The next proposition shows that the test statistic $T_{i,j}$ and the auxiliary statistic $A_{i,j}$ are asymptotically independent under the null hypothesis. Define 
\begin{eqnarray*}
a_{i,j} = \frac{s_{1,i,j}+\kappa_{i,j}s_{2,i,j}}{\left\{ \Var(S_{1,l,i,j})+\kappa_{i,j}^2\Var(S_{1,l,i,j}) \right\}^{1/2}}, \;\; \textrm{ where } \;
\kappa_{i,j} = \frac{ n_2\Var(S_{1,l,i,j}) } { n_1\Var(S_{2,l,i,j}) }.
\end{eqnarray*}

\begin{proposition}\label{asymp.indep}
Suppose (C1) holds. For any constants $M>0$ and $C>0$, we have
\begin{eqnarray*}
\pr_{H_{0,i,j}}\left( |T_{i,j} |\geq h, |A_{i,j}|\geq \lambda \right) = \{ 1+o(1) \} G(h) \pr\left( |N(0,1)+a_{i,j}|\geq \lambda \right) + O(q^{-M}),
\end{eqnarray*}
uniformly for $0\leq h\leq C\sqrt{\log q}$, $0\leq \lambda\leq C\sqrt{\log q}$, and $1\leq i<j\leq p$, with $G(h)=2\{1-\Phi(h)\}$. Furthermore, for all $0\leq k\leq CN$ with an integer constant $N$,
\begin{eqnarray*}
\pr_{H_{0,i,j}}\left( |T_{i,j} |\geq h, |A_{i,j}|< \lambda_k \right) = \{1+o(1)\} G(h) \pr\left( |N(0,1)+a_{i,j}|< \lambda_k \right) + O(q^{-M}),
\end{eqnarray*}
uniformly for $0\leq h\leq C\sqrt{\log q}$ and $1\leq i<j\leq p$, where $\lambda_k=(k/N)\sqrt{\log q}$.
\end{proposition}

\subsection{Power enhanced simultaneous test}\label{GAP.sec}

Based on $(T_{i,j}, A_{i,j})$, we now modify the simultaneous testing procedure of Algorithm \ref{proc1}. We first describe the main idea. We next summarize the modified testing procedure in Algorithm \ref{proc2}. Finally, we discuss some specific choices of the key parameters of the algorithm. 

Since there are totally $q = p(p-1)/2$ tests to carry out simultaneously, we rearrange the pairs of $\{(T_{i,j}, A_{i,j}), 1\leq i<j\leq p\}$ into $\{(T_i,A_i),i=1,\ldots,q\}$. After obtaining all the $p$-values, $p_i=2\{ 1-\Phi(|T_i|) \}$, from Algorithm \ref{proc1}, our basic idea is to adjust those $p$-values by $p_i^w=\min\{p_i/w_i,1\}$, with $w_i$ being the adjusting weights, $i=1,\ldots, q$. We utilize the auxiliary statistics $A_i$ to help compute the adjusting weights $w_i$, by groups. Specifically, we consider a set of grid values, $\cJ = \{(C_1N-1) \sqrt{\log q}/N, C_1 \sqrt{\log q}, \ldots, (C_2N-1) \sqrt{\log q} / N, C_2 \sqrt{\log q} \}$, where $C_1$, $C_2$ and $N$ are some pre-specified constants. We divide the index set $\{1, \ldots, q\}$ into $K$ groups according to the auxiliary statistics $(A_1, \ldots, A_q)$. As an example, we take $K = 3$. That is, we choose two grid points $\cJ_{K} = \{\lambda_1, \lambda_2\}$ in $\cJ$, and obtain $K=3$ groups of indices, $\cG_1 = \{i \, : \, 1 \leq i \leq q, \,  -\infty < A_i \leq \lambda_1\}$, $\cG_2 = \{i \, : \, 1 \leq i \leq q, \,  \lambda_1 < A_i \leq \lambda_2\}$, and $\cG_3 = \{i \, : \, 1 \leq i \leq q, \,  \lambda_2 < A_i \leq \infty \}$. For each group $\cG_k$, we compute its cardinality, $q_k = |\cG_k|$. We also estimate the proportion, $\pi_k$, of alternatives in $\cG_k$, $k = 1, \ldots, K$. To do so, we employ the method of \citet{schweder1982plots} and \cite{storey2002direct} to obtain an estimate $\tilde\pi_k$ first, then stablize it by $\hat\pi_k=(\epsilon\vee \tilde\pi_k)\wedge(1-\epsilon)$, where $\epsilon$ is a small positive number; we set $\epsilon = 10^{-5}$. Then for all the indices in $\cG_k$, we compute the group-wise adjusting weight:  
\begin{eqnarray}\label{weight}
w_i=\left( \sum_{k=1}^K \frac{ q_k \hat\pi_k}{1 - \hat\pi_k} \right)^{-1} \frac{q \hat\pi_k}{(1- \hat\pi_k)}, \quad i \in \cG_k, \; 1\leq k\leq K. 
\end{eqnarray}
This idea of adjusting the weights $w_i$ by groups is motivated by our intuition discussed in Section \ref{intuition.sec}. After obtaining the weights, we adjust the $p$-values and apply the Benjamini-Hochberg procedure \citep[BH]{BH1995} to the adjusted $p$-values $p_i^w$. Finally, we search all possible choices of $\cJ_{K}$ among $\cJ$, and find the one that yields the largest number of rejections. We apply the BH procedure again to the adjusted $p$-values under this choice of $\cJ_{K}$ to obtain the final adjusted rejection region. We summarize this modified simultaneous testing procedure in Algorithm \ref{proc2}.

\begin{algorithm}[t!]
\caption{Adjusted simultaneous inference with FDR control and power enhancement.}
\label{proc2}
\begin{algorithmic}
\item Step 1: Initialization:
\vspace{-0.1in}
\begin{enumerate}[]
\item[] Step 1.1: Compute the test statistics and auxiliary statistics $\{(T_i,A_i),i=1,\ldots,q\}$.
\item[] Step 1.2: Compute the $p$-values: $p_i=2\{ 1-\Phi(|T_i|) \}$, $i=1,\ldots, q$.
\item[] Step 1.3: Input the pre-specified constants $K$, $C_1$, $C_2$ and $N$. 
\item[] Step 1.4: Compute the grid set:
\vspace{-0.1in}
\[ 
\cJ = \left\{ (C_1N-1) \sqrt{\log q}/N, C_1 \sqrt{\log q}, \ldots, (C_2N-1) \sqrt{\log q} / N, C_2 \sqrt{\log q} \right \}. 
\]
\end{enumerate} 
\vspace{-0.1in}

\item Step 2: For each $\cJ_{K} = \{\lambda_1, \ldots, \lambda_{K-1}\} $ in $\cJ$, and $\lambda_0 = -\infty, \lambda_K = \infty$: 
\begin{enumerate}[]
\item[] Step 2.1: Construct $\cG_k = \{i \, : \, 1 \leq i \leq q, \, \lambda_{k-1} < A_i \leq \lambda_k\}$, $1\leq k\leq K$. 
\item[] Step 2.2: For each $\cG_k$, compute the cardinality, $q_k = |\cG_k|$.
\item[] Step 2:3: For each $\cG_k$, estimate the proportion, $\hat\pi_k$, of alternatives in $\cG_k$.
\item[] Step 2.4: Compute the adjusting weights $w_i$, $i=1,\ldots, q$, according to \eqref{weight}.
\item[] Step 2.5: Adjust the $p$-values: $p_i^w=\min\{p_i/w_i,1\}$, $i=1,\ldots, q$.
\item[] Step 2.6: Apply the BH procedure, and record the total number of rejections. 
\end{enumerate}

\item Step 3: Obtain the adjusted rejection region: 
\begin{enumerate}[]
\item[] Step 3.1: Choose $\cJ_{K}$ that yields the largest number of rejections.
\item[] Step 3.2: Compute the corresponding adjusted $p$-values: $p_{i}^{w}$, $1\leq i \leq  q$. 
\item[] Step 3.3: Reorder all the adjusted $p$-values: $p_{(1)}^{w} \leq \ldots \leq p_{(q)}^{w}$.
\item[] Step 3.4: Output the rejection region $\{ i \, : \, i < \hat\tau \}$, where $\hat\tau = \max\{i \, : \, p^w_{(i)} \leq \alpha i/q\}$.
\end{enumerate}
\end{algorithmic}
\end{algorithm}

We discuss some specific choices of the parameters in Algorithm \ref{proc2}. First, the number of groups $K$ is usually set at $K=3$. As shown in \citet{Xia2018GAP}, when $K\geq 4$, there is little additional power gain, but a more expensive computation. Second, the constants $C_1$ and $C_2$ can be chosen so that $C_1\sqrt{\log q}$ is equal to the smallest value of the auxiliary statistics and $C_2\sqrt{\log q}$ is equal to the largest value of the auxiliary statistics. If the absolute values of the smallest and largest auxiliary statistics exceed $16\sqrt{\log q}$, we truncate at $C_1=-16$ and $C_2=16$ to stabilize and expedite the computation. We note here that, if the network data are non-negative, such as the binary and poisson network data, then both $C_1$ and $C_2$ are non-negative. By contrast, in \citet{Xia2018GAP}, $C_1$ and $C_2$ were fixed at $-4$ and $4$. Finally, $N$ can be any integer for the theoretical validity. Numerically, a larger value of $N$ implies a more precise grid search, but at the cost of a heavier computational burden. We choose $N$ such that the gap between two adjacent grid points, $(\log p)^{1/2}/N$, equals $0.1$ approximately.

\subsection{FDR control and power enhancement}\label{GAPtheory.sec}

We next show that the modified inference of Algorithm \ref{proc2} is asymptotically more powerful than Algorithm \ref{proc1}, meanwhile it can still asymptotically control the false discovery. 

Denote $\{p_i^w: 1 \leq  i \leq q \}$ the adjusted $p$-values from Algorithm \ref{proc2}, and $\{p_{(i)}^w: 1 \leq  i \leq q \}$ the ordered adjusted $p$-values. The corresponding adjusted FDP is:
\vspace{-0.1in}
\begin{eqnarray*}
\textrm{FDP}_{\textrm{adj}} = \frac{\sum_{i\in\cH_0} I\left\{p^w_i\leq p^w_{(\hat{\tau})}\right\}}{\sum_{i=1}^q I\left\{p^w_i\leq p^w_{(\hat{\tau})}\right\} \vee 1}, 
\end{eqnarray*}
where $\hat{\tau}$ is the cutoff obtained from Step 3.4 of Algorithm \ref{proc2}, and $I(\cdot)$ is the indicator function. Accordingly, $\textrm{FDR}_{\textrm{adj}} = \ep(\textrm{FDP}_{\textrm{adj}})$. The next theorem shows that the modified procedure can still control FDR and FDP asymptotically.

\begin{theorem}
\label{FDRGAP}
Suppose (A2), (B1)-(B3), and (C1) hold. Suppose $\log q=o(n^{1/C})$ for some constant $C>5$. Then, 
\begin{eqnarray*}
\lim_{(n_1,n_2,q)\rightarrow \infty}\frac{\textrm{FDR}_{\textrm{adj}}}{\alpha q_0/q}=1,
\quad \textrm{and} \quad
\frac{\textrm{FDP}_{\textrm{adj}}}{\alpha q_0/q} \rightarrow 1 \; \textrm{ in probability, as } n_1,n_2,q \rightarrow \infty.
\end{eqnarray*} 
\end{theorem}

Next, denote the power of the testing procedures of Algorithms \ref{proc1} and \ref{proc2} by $\Psi$ and $\Psi_{\textrm{adj}}$, respectively. That is, 
\begin{eqnarray*}
\Psi = \ep\left\{\frac{\sum_{(i,j)\in\cH_1}I(|T_{i,j}|\geq \hat h)}{|\cH_1|}\right\}, \;
\Psi_{\textrm{adj}} = \ep\left[\frac{\sum_{i\in\cH_1}I\left\{ p^w_i\leq p^w_{(\hat{\tau})} \right\}}{|\cH_1|}\right].
\end{eqnarray*}
Then the next theorem shows that, by incorporating the auxiliary statistics $A_{i,j}$, the modified simultaneous testing procedure of Algorithm \ref{proc2} is asymptotically more powerful than Algorithm \ref{proc1}, which is solely based on the test statistics $T_{i,j}$.

\begin{theorem}
\label{power.compare}
Suppose the same conditions in Theorem \ref{FDRGAP} hold. Then, 
\[
\Psi_{adj} \geq \Psi + o(1), \;\; \textrm{ as } \;\; q \rightarrow \infty. 
\]
\end{theorem}

\subsection{Comparison to GAP}\label{compare.sec}

Although motivated by the GAP method of \citet{Xia2018GAP}, our power enhancement procedure is also considerably different from GAP. While \citet{Xia2018GAP} tackled the problem of mean comparison of vector-valued samples, we target the problem of network mean comparison. This leads to a different set of test and auxiliary statistics, but a number of additional intrinsic differences as well. 

First, the two methods impose different assumptions. A key requirement for GAP to enhance the power is that the parameters of interest from each group are individually sparse. In our setup, however, the parameters may all be non-negative. For instance, in a binary network or a count network, all the entries of both means $\s_1$ and $\s_2$ are usually non-negative. As such, the means may not be individually sparse.  Our procedure instead only requires the difference of the two means $\s_1 - \s_2$ is sparse, which reasonably holds and is often imposed in numerous applications including brain connectivity analysis \citep{ZhuLi2018}.  

Second, the two methods differ in terms of the range of the auxiliary statistics that contribute most to the power enhancement. Consider the case when $K = 3$. In \citet{Xia2018GAP}, since both means are assumed to be individually sparse, the tests that are more likely to be adjusted and rejected are those with the corresponding auxiliary statistics either being negative and small, or positive and large. That is, the power enhancement hinges more on those tests in $\cG_1$ and $\cG_3$ with small or large auxiliary statistics. However, in our setup, the individual means $\s_1$ and $\s_2$ can both be dense and their entries are all positive. Instead we only assume that $\s_1-\s_2$ is sparse. Take a binary brain connectivity network as an example. The observed networks are often sparse, in that most links are zero, since it is known that brain connections are energy consuming and biological units tend to minimize energy-consuming activities \citep{Raichle2002, Bullmore2009}. This translates to small connection probabilities for most entries of $\s_1$ and $\s_2$, while all these probabilities are positive. Moreover, the difference of the means between the two populations is often sparse, which translates to equal connection probabilities for most entries of $\s_1$ and $\s_2$, or equivalently zero difference for most entries of of $\s_1 - \s_2$. This is similar to the toy example we discuss in Section \ref{intuition.sec}. For such cases, as a consequence of Algorithm \ref{proc2}, the tests whose corresponding auxiliary statistics are too small or too large would be adjusted so that they are less likely to be rejected. Instead, those tests whose auxiliary statistics are in between would be adjusted so that they are more likely to be rejected. In other words, the power enhancement in our setup may hinge more on $\cG_2$, rather than $\cG_1$ and $\cG_3$.  

Third, due to the above difference, the grid construction in Step 1.4 of Algorithm \ref{proc2} is noticeably different from that of GAP in \citet{Xia2018GAP}. Specifically, in \citet{Xia2018GAP}, to ensure the inclusion of important locations in $\cG_1$ and $\cG_3$, the constants $C_1$ and $C_2$ can be simply fixed at $-4$ and $4$, respectively, so that the upper bound of those small negative auxiliary statistics and lower bound of those large positive auxiliary statistics can be attained in the grid $\cJ$. By contrast, for our problem, the upper bound of the auxiliary statistics in the union support $\cI$ can go beyond the bound in \citet{Xia2018GAP}, i.e., $4\sqrt{\log q}$, and the lower bound of the auxiliary statistics in the union support can be non-negative. Since the tests in $\mathcal G_2$ are more likely to be adjusted and rejected, we need to do a more thorough grid construction and choose the constants $C_1$ and $C_2$ based on the smallest and largest values of the auxiliary statistics as described in Section \ref{GAP.sec}.

\section{Numerical Analysis}
\label{numerical.sec} 

We first study and compare the finite-sample performance of the two simultaneous inference procedures, Algorithms \ref{proc1} and \ref{proc2}, through simulations. We then illustrate our method with a brain structural connectivity analysis.

\subsection{Simulations}\label{simulation.sec}

We consider a $68 \times 68$ network, which is of the same dimension as the real data in Section \ref{data.sec}. This results in $q=68(68-1)/2 = 2278$ hypotheses to test simultaneously. We consider three network data distributions, three sparsity levels, and two different sample sizes. More specifically, we consider Bernoulli, Bernoulli mixture, and transformed Wishart distributions. 
\begin{itemize}
\item
{\bf Bernoulli: } Select the sets $\cM_{d,1}$ and $\cM_0$ from $q$ hypotheses, uniformly and randomly, with $|\cM_{d,1}|=|\cM_0|=k_q/2$, $d=1,2$. Here $k_q$ is a parameter that controls the sparsity level, and is defined later. Let $\cM_{d} = \cM_{d,1}\cup\cM_0$. For $(i,j)\notin \cM_d$, generate $S_{d,l,i,j} \sim \text{Bernoulli}(1,0.3)$. For $(i,j)\in \cM_d$, generate $S_{d,l,i,j} \sim \text{Bernoulli}(1,r_{d,i,j})$, where $r_{1,i,j}$ is equal to 0.5 with probability 0.1, and is equal to 0.8 otherwise, whereas $r_{2,i,j}$ is equal to 0.8 with probability 0.1, and is equal to 0.5 otherwise.

\item
{\bf Bernoulli mixture: } Generate $\cM_{d}$ in the same way as before. Generate $S_{d,l,i,j} \sim \text{Bernoulli}(1,r_{d,i,j})$, where $r_{d,i,j} = \pi_{i,j}*r_{d,1,i,j} + (1-\pi_{i,j})*r_{d,2,i,j}$, with $\pi_{i,j} \sim \text{Uniform}(0,1)$. For $(i,j)\notin \cM_d$, $r_{d,1,i,j}=r_{d,2,i,j}=0.3$, $d=1,2$. For $(i,j)\in \cM_d$, $r_{1,1,i,j}$ is equal to 0.5 with probability 0.1, and is equal to 0.7 otherwise, whereas $r_{2,1,i,j}$ is equal to 0.7 with probability 0.1, and is equal to 0.5 otherwise, and $r_{d,2,i,j} = r_{d,1,i,j} + 0.2$.

\item
{\bf Wishart with logarithm transformation: } Select the sets $\cM_{d,1}$ and $\cM_0$ from $q$ hypotheses, uniformly and randomly, with $|\cM_{d,1}|=k_q/4$, and $|\cM_0|=3k_q/4$, $d=1,2$. Let $\cM_{d} = \cM_{d,1}\cup\cM_0$. Generate $\Sigma_d$ such that $\Sigma'_{d,i,j} = \text{Uniform}(3,5)$ if $(i,j)\in \cM_d$ and $\Sigma'_{d,i,j} = 0$ otherwise. Let $\Sigma'_{d,j,i} = \Sigma'_{d,i,j}$ and $\Sigma_d = \Sigma'_d + \{ |\lambda_{\min}(\Sigma'_d)| + 0.5 \} I$, where $I$ is the identify matrix. Generate $S_{d,l}'\sim \text{Wishart}(m^{-1}\Sigma_{d},m)$, with $m=100$, and $S_{d,l} = \log[ \text{round}\{ \exp(S_{d,l}')\} ]$, where round$(\cdot)$ rounds a number to the nearest integer. 
\end{itemize}

\noindent
The sparsity level between the two network population means is controlled by the parameter $k_q$, and we consider three sparsity levels, $k_q = 0.2q, 0.15q$ and $0.1q$. We also examine two sample sizes, $n_1=n_2=100$ and $n_1=n_2=25$, and the latter mimics the real data setting where the sample size is very limited.

\begin{table}[p]\small\addtolength{\tabcolsep}{-4pt}
\begin{center}
\caption{\small The empirical FDR and empirical power, in percentage, for Algorithms \ref{proc1} and \ref{proc2} based on 100 data replications. The significance level is $\alpha=5\%$.}
\label{table-sim}
\begin{tabular}{ccccccc} \toprule
 &\multicolumn{3}{c}{$n_1=n_2=100$} & \multicolumn{3}{c}{$n_1=n_2=25$} \\ \cmidrule(r){2-4} \cmidrule(r){5-7}
\multicolumn{1}{c}{} & 
\multicolumn{1}{c}{\mbox{ } $0.2q$ \mbox{ }} & \multicolumn{1}{c}{\mbox{ } $0.15q$ \mbox{ }} & \multicolumn{1}{c}{\mbox{ } $0.1q$ \mbox{ }} &
\multicolumn{1}{c}{\mbox{ } $0.2q$ \mbox{ }} & \multicolumn{1}{c}{\mbox{ } $0.15q$ \mbox{ }} & \multicolumn{1}{c}{\mbox{ } $0.1q$ \mbox{ }} \\ \midrule
Bernoulli  &\multicolumn{6}{c}{Empirical FDR} \\ \cmidrule(r){2-7}
Algorithm 1        & 4.0 & 4.5 & 4.9 & 5.8 & 6.4 & 7.3\\
Algorithm 2	 & 2.9 & 2.8 & 3.5 & 3.9 & 5.0 & 5.9\\ \midrule
 &\multicolumn{6}{c}{Empirical power}\\ \cmidrule(r){2-7}
Algorithm 1        & 88.1 & 86.7 & 84.6 & 44.6 & 42.2 &40.1 \\
Algorithm 2	 & 91.6 & 91.5 & 90.7 & 55.3 & 54.7 &53.2 \\ 
\bottomrule      
Bernoulli mixture &\multicolumn{6}{c}{Empirical FDR} \\ \cmidrule(r){2-7}
Algorithm 1        & 4.1 & 4.4 & 4.6 & 5.9 & 5.8 & 7.5\\
Algorithm 2	 & 1.5 & 1.6 & 2.4 & 3.5 & 4.1 & 5.4\\ \midrule
 &\multicolumn{6}{c}{Empirical power}\\ \cmidrule(r){2-7}
Algorithm 1        & 89.8 & 88.6 & 87.6 & 42.3 & 41.4 & 41.1\\
Algorithm 2	 & 95.7 & 95.8 & 95.6 & 54.6 & 54.7 & 54.5\\
\bottomrule      
Transformed Wishart &\multicolumn{6}{c}{Empirical FDR} \\ \cmidrule(r){2-7}
Algorithm 1        & 4.4 & 4.9 & 5.4 & 4.9 & 5.5 &6.2 \\
Algorithm 2	& 2.3 & 2.5 & 2.7 & 3.0 & 3.8 &4.2 \\ \midrule
 &\multicolumn{6}{c}{Empirical power}\\ \cmidrule(r){2-7}
Algorithm 1        & 52.2 & 55.1 & 59.3 & 37.3 & 39.4 &42.0 \\
Algorithm 2	& 60.1 & 64.4 & 69.4 & 41.6 & 44.7 &49.8 \\
\bottomrule      
\end{tabular}
\end{center}
\end{table}

We apply both Algorithms \ref{proc1} and \ref{proc2}, and set the nominal level at $\alpha=0.05$. We report the empirical FDR and power, both in percentage, based on 100 replications in Table \ref{table-sim}. It is seen that, in all cases, the empirical FDRs are generally controlled under the nominal level by both algorithms. Algorithm \ref{proc2} is slightly more conservative than Algorithm \ref{proc1}, which is mainly due to the normalization step of the weight calculation as shown in \eqref{weight}. A similar phenomenon has also been observed in \citet{Xia2018GAP}. For the empirical power, it is seen that Algorithm \ref{proc2} has a clear power improvement over Algorithm \ref{proc1}, thanks to its utilization of the auxiliary information. Furthermore, the performance under the varying sample size confirms the power enhancement of Algorithm \ref{proc2} as theoretically revealed in Section \ref{GAPtheory.sec}. We also observe that the power gain becomes more substantial when the true difference $\s_1-\s_2$ becomes more sparse, which agrees with our intuition explained in Section \ref{intuition.sec}.

\subsection{Structural connectivity analysis}\label{data.sec}

We illustrate our method with a brain structural connectivity analysis of the KKI-42 dataset, which is available at \texttt{http://openconnecto.me/data/public/MR/archive/}, and its detailed description can be found in \citet{landman2011multi}. The data consists of 21 subjects with no history of neurological conditions, aging from 22 to 61 years old. Each subject received two resting-state diffusion tensor images (DTI) under a scan-rescan imaging session. For simplicity, we treat the data as if those images were from independent samples, which is common for the analysis of this dataset \citep{wang2017common}. It results in a total sample size of $42$ for this study. Brain regions are constructed following the Desikan Atlas \citep{desikan2006automated}, leading to $p=68$ regions equally divided in the left and right hemispheres. DTI is a magnetic resonance imaging technique that measures the diffusion of water molecules to map white matter tractography in the brain. Each DTI image was preprocessed, and was summarized in the form of a $68 \times 68$ network, where the edges record the total number of white matter fibers between the pair of nodes. It is also equally common to focus on the form of a binary network, where the edges become the binary indicators of presence of absence of white matter fibers \citep{wang2017common}. See \citet{Zhang2018} for more details on the construction of a brain structural network from DTI images. We partition the subjects into two age groups, the ones whose are younger than 30 years, and the ones who are 30 or older. Age 30 is a transition period, usually known as the ``age 30 transition", when the first phase of early adulthood comes to a close, and the basis for the next life structure is formed. Moreover, this partition yields about the same number of subjects for each group, with $n_1 = 22$ for the younger-than-30 age group, and $n_2 = 20$ for the older-than-30 age group. We study the age-related difference in structural connectivity patterns, which is of universal interest, as aging is the main risk factor for progressive loss of either structure or function of brain neurons \citep{MorrisonHof1997}.

\begin{figure}[t!]
\centering{
\caption{Differentiating links and the associated brain regions found by the proposed power-enhanced multiple testing procedure for the KKI-42 dataset of the binary network. \label{fig:kki-binary}}
\begin{tabular}{cc}
\vspace{-0.07in}
axial view & coronal view \\
\includegraphics[height=2.3in,width=3.2in]{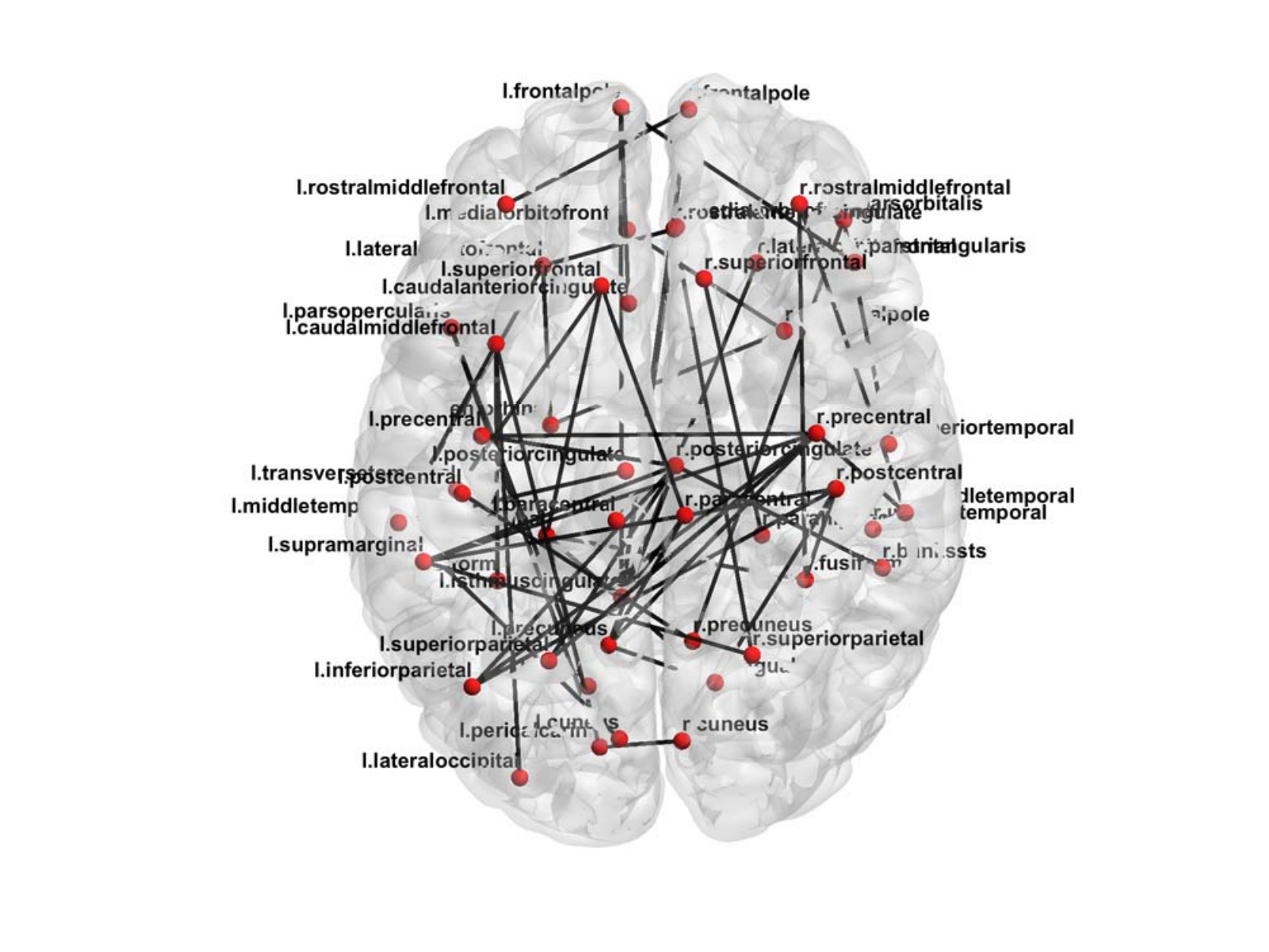} & 
\includegraphics[height=2.2in,width=2.8in]{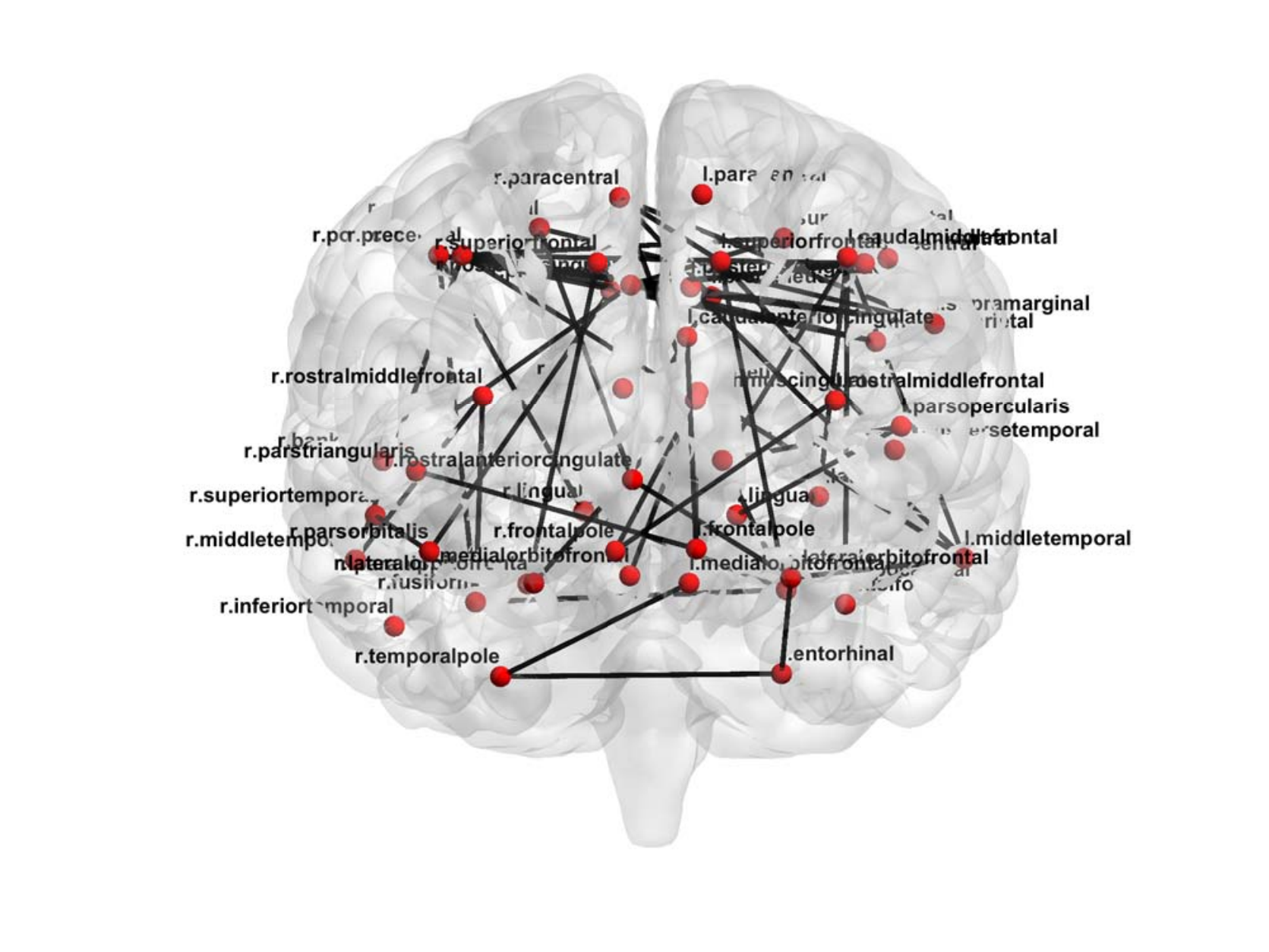} \\
\vspace{-0.03in}
left sagittal view & right sagittal view \\
\includegraphics[height=1.9in,width=2.7in]{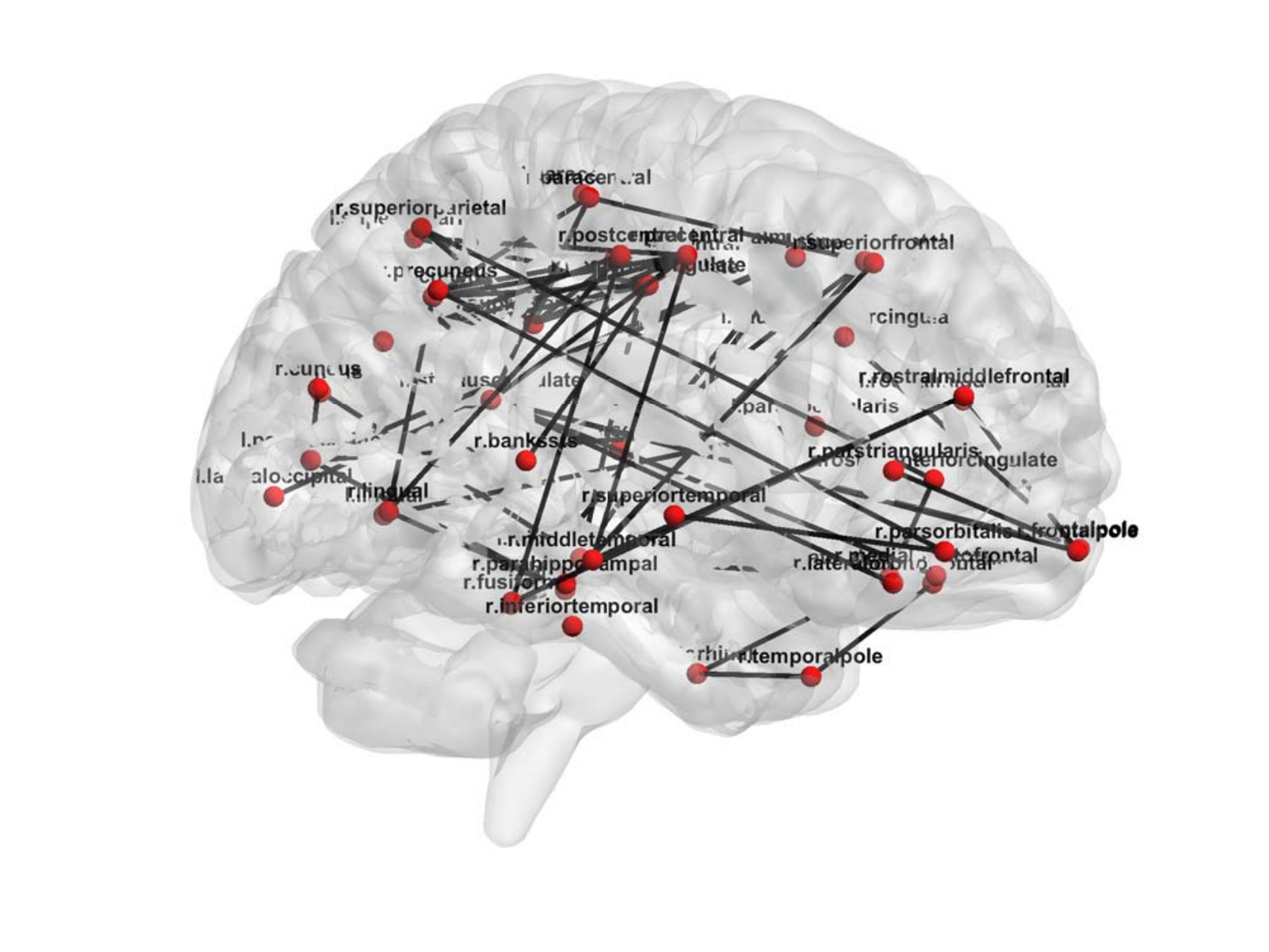} & 
\includegraphics[height=1.9in,width=2.7in]{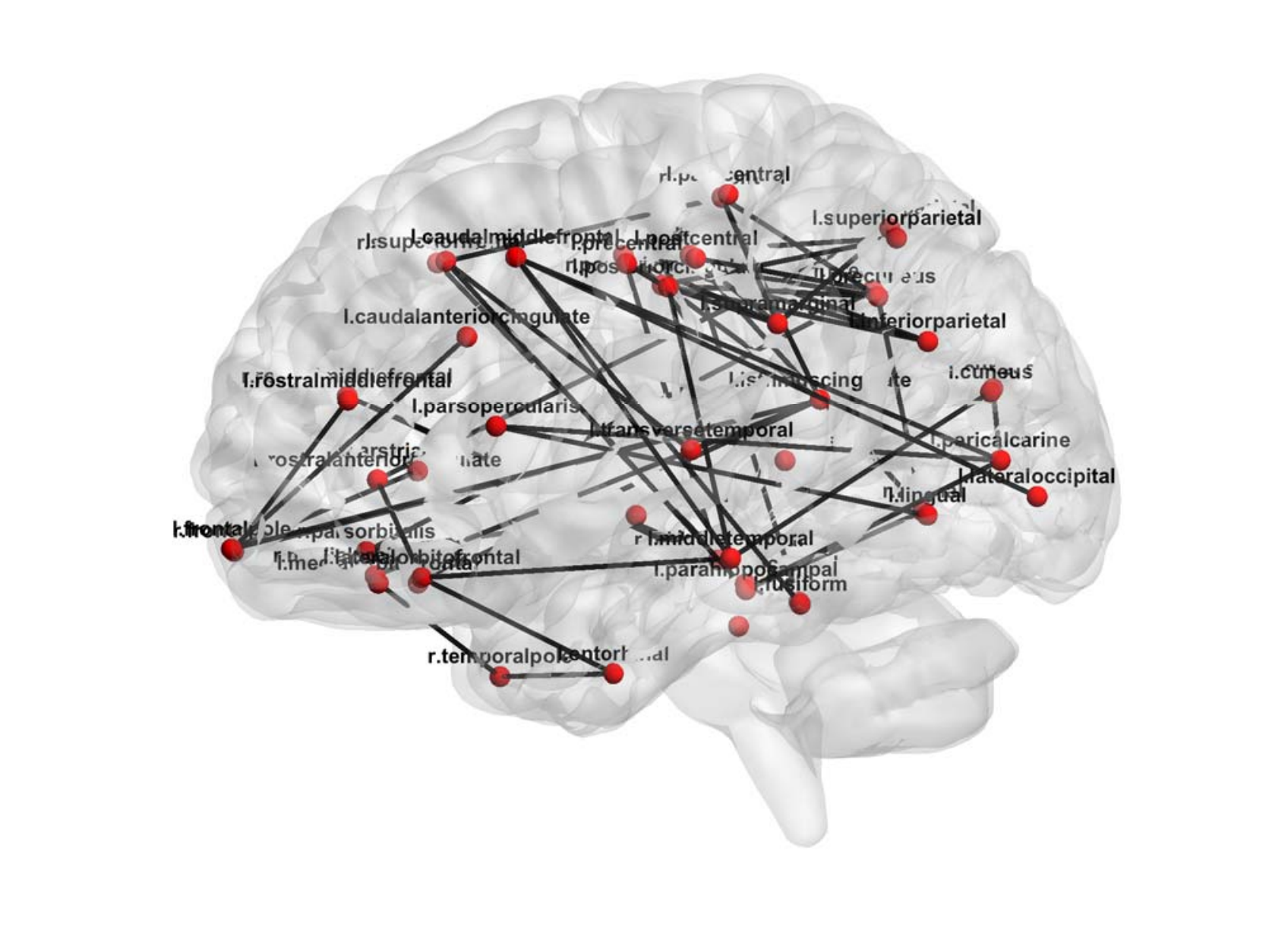} \\
\end{tabular}
}
\end{figure}

\begin{figure}[t!]
\centering{
\caption{Differentiating links and the associated brain regions found by the proposed power-enhanced multiple testing procedure for the KKI-42 dataset of the count network. \label{fig:kki-count}}
\begin{tabular}{cc}
\vspace{-0.07in}
axial view & coronal view \\
\includegraphics[height=2.3in,width=3.2in]{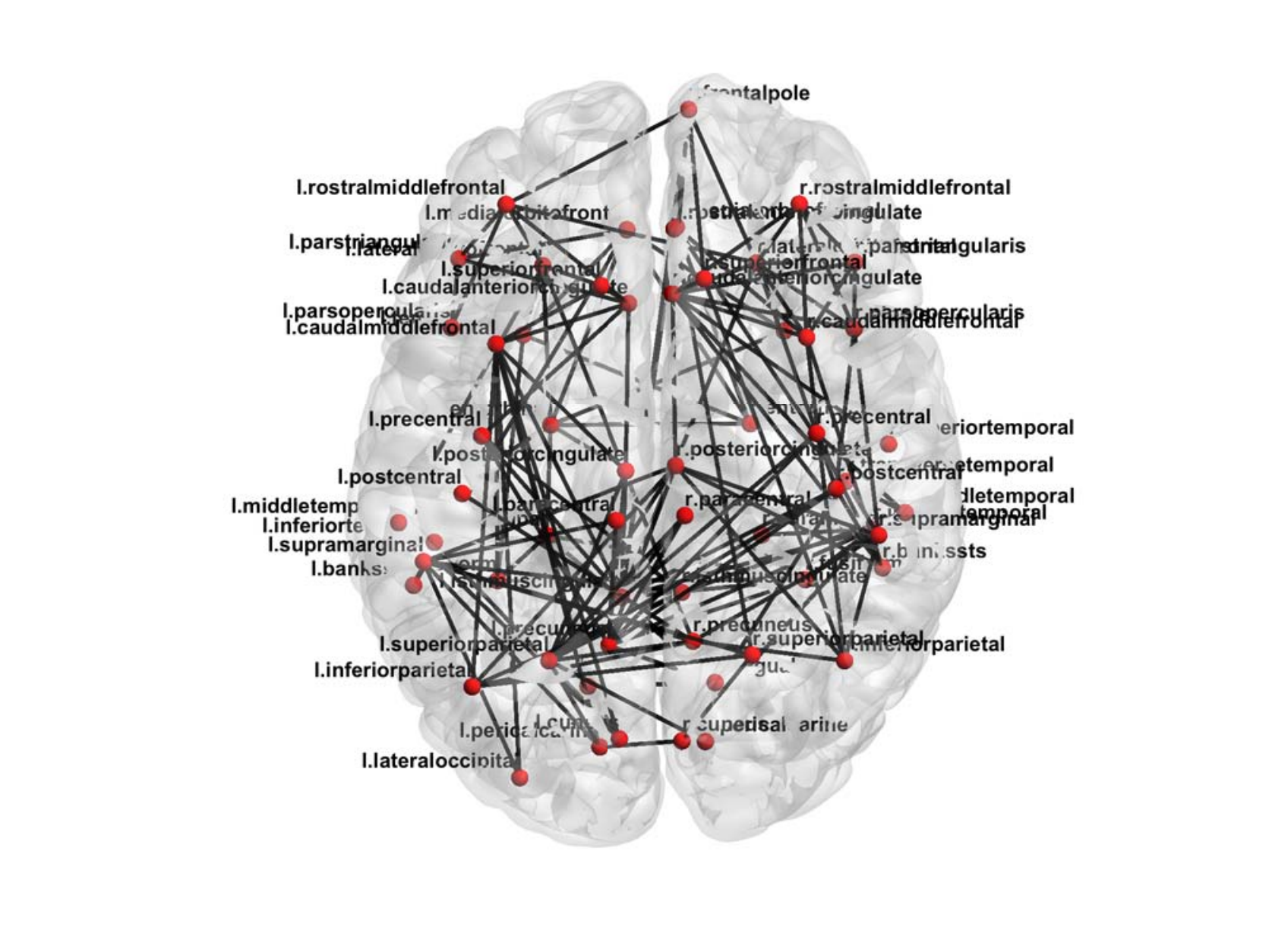} & 
\includegraphics[height=2.2in,width=2.8in]{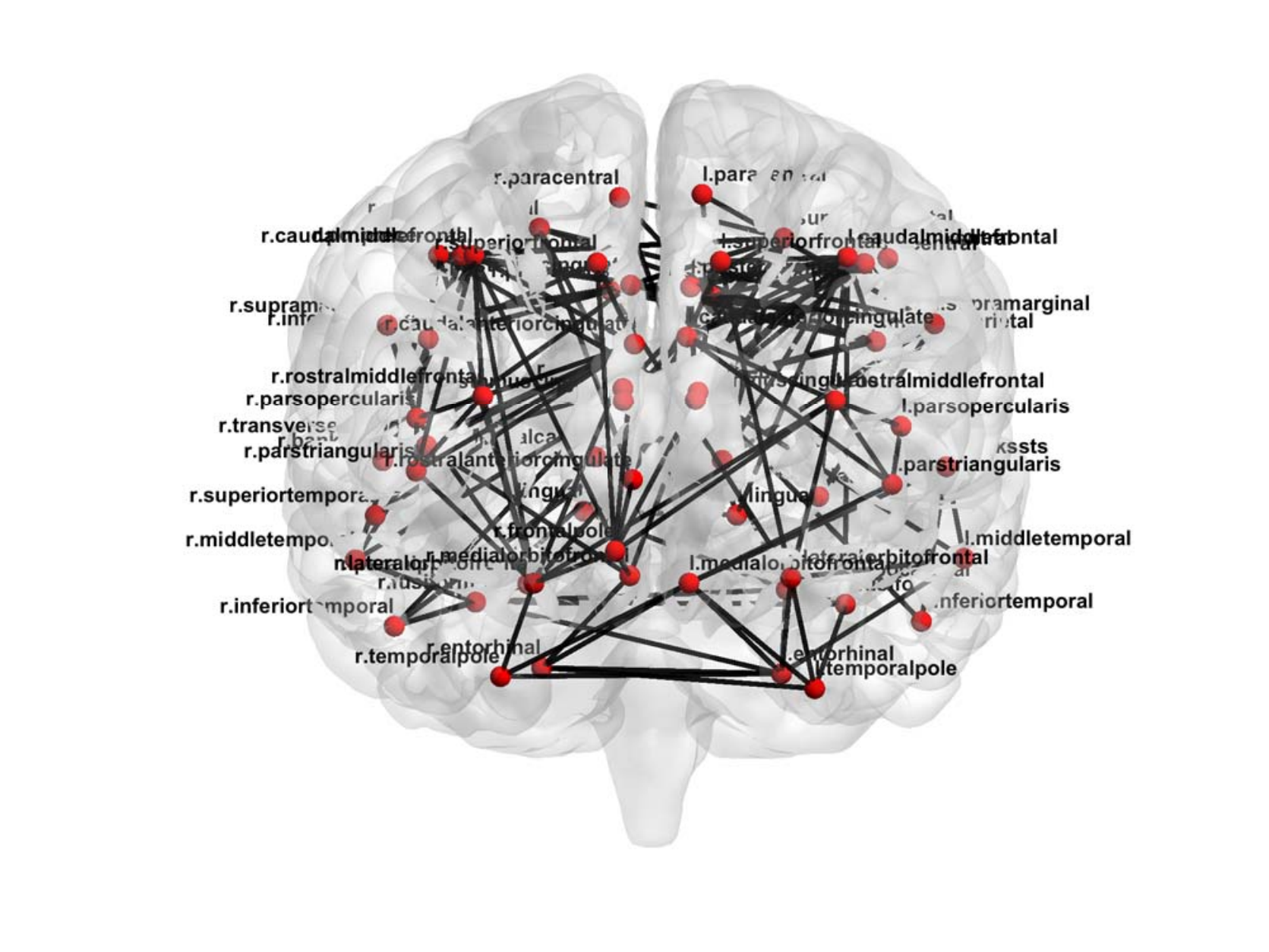} \\
\vspace{-0.03in}
left sagittal view & right sagittal view \\
\includegraphics[height=1.9in,width=2.7in]{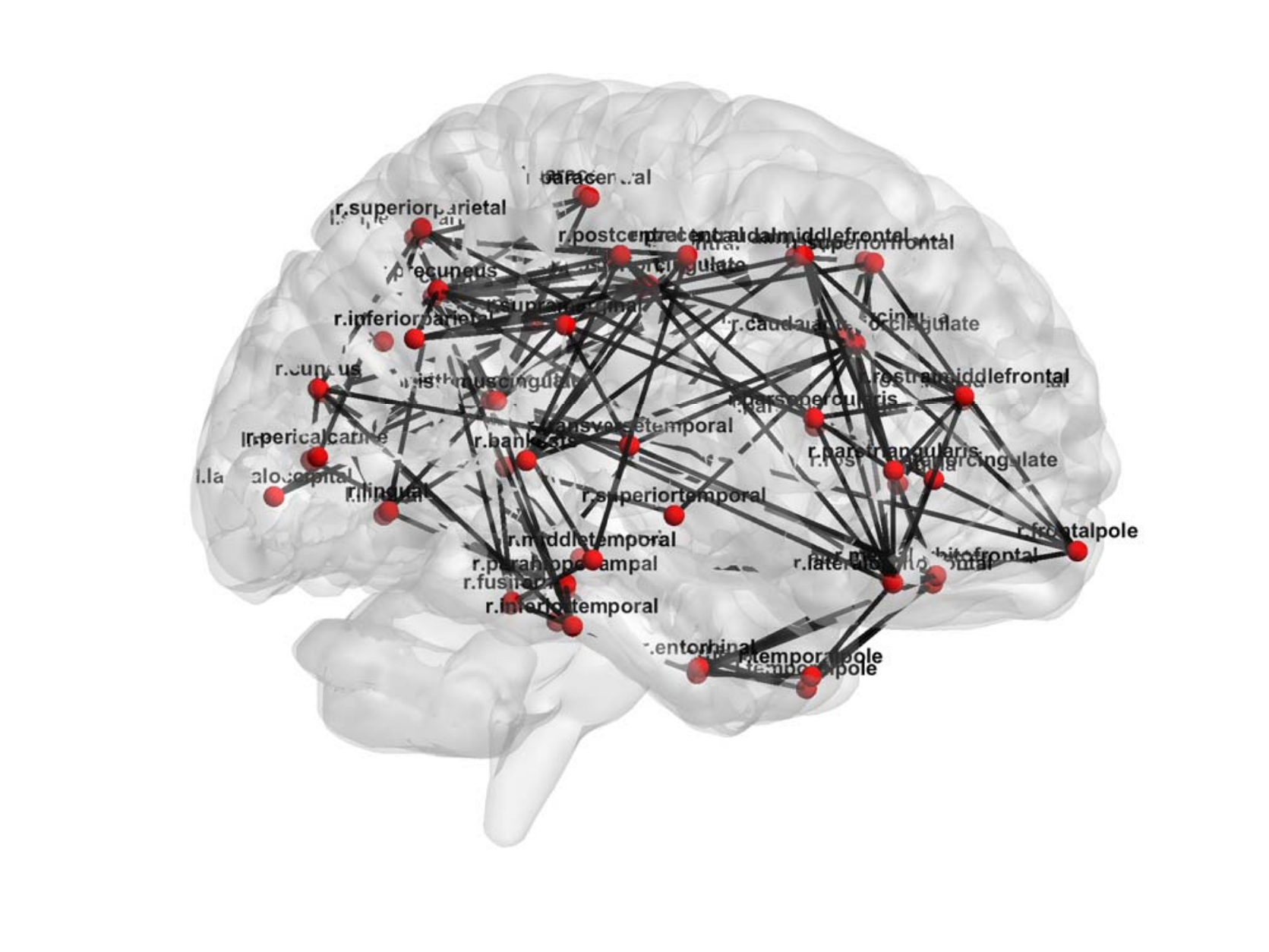} & 
\includegraphics[height=1.9in,width=2.7in]{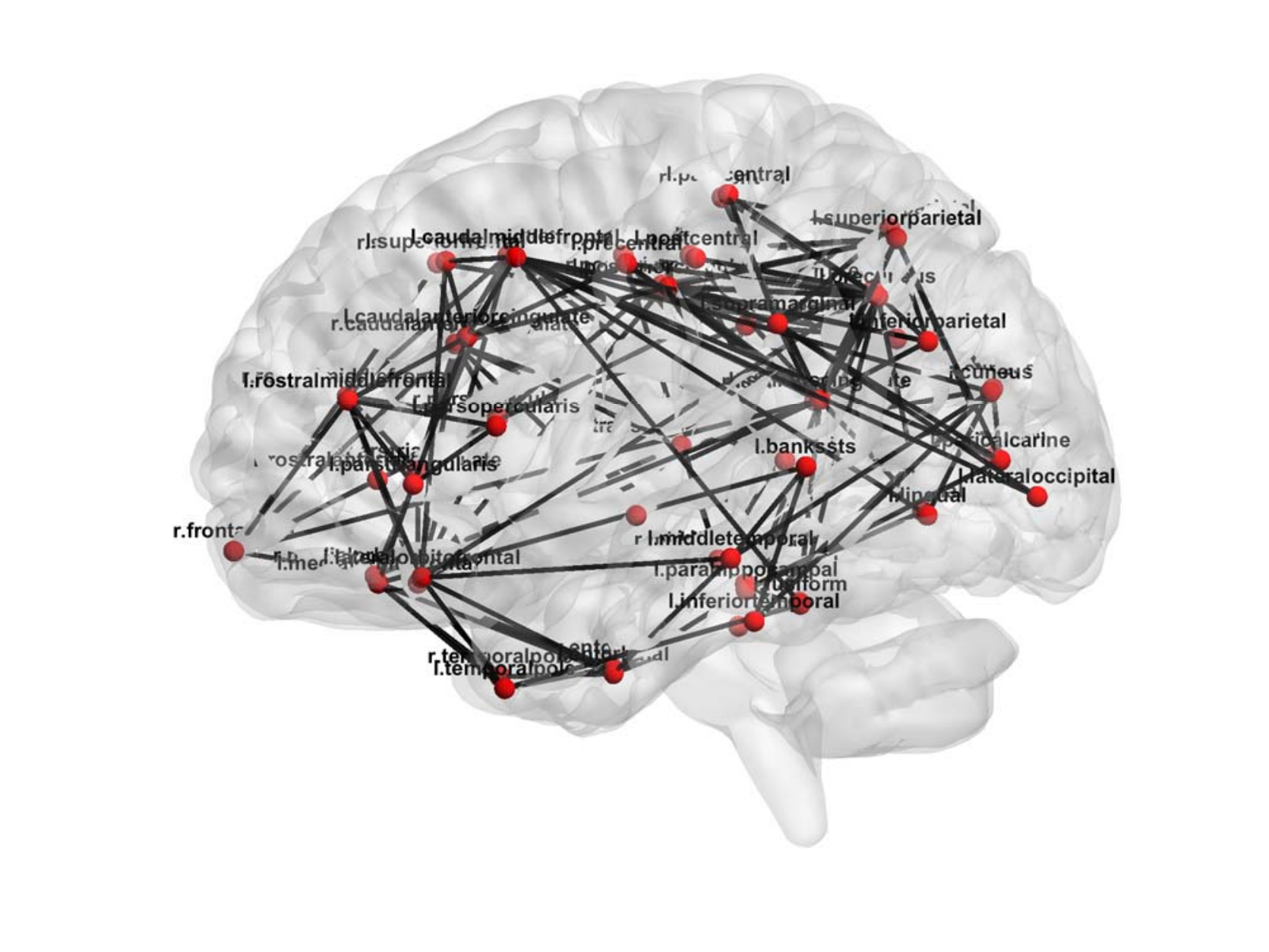} \\
\end{tabular}
}
\end{figure}

We applied the power-enhanced test to this dataset, first the binary network, then the count network with a logarithm transformation. We set the significance level at 0.05. For the binary network, out of the total of 2278 links, Algorithm \ref{proc1} identified 38 significantly different links, whereas the power-enhanced Algorithm \ref{proc2} identified 62 links, including all 38 linked found by Algorithm \ref{proc1} plus 24 additional links. This agrees with both our theory and the simulations, in that Algorithm \ref{proc2} is usually able to recognize more significant links than Algorithm \ref{proc1}. Figure \ref{fig:kki-binary} plots all the identified links by Algorithm \ref{proc2} and the associated brain regions, visualized using the BrainNet Viewer \citep{Xia2013}. Table \ref{tab:binary-addition} records the 24 additional links found by Algorithm \ref{proc2}. For the count network, out of the total of 2278 links, Algorithm \ref{proc1} identified 140 significantly different links, whereas Algorithm \ref{proc2} identified 177 links. Among the 140 links found by Algorithm \ref{proc1}, 125 links were also found by Algorithm \ref{proc2}. Moreover, Algorithm \ref{proc2} identified 52 additional links. Figure  \ref{fig:kki-count} shows all the identified links by Algorithm \ref{proc2}. Those additional links found by Algorithm \ref{proc2} capture the smaller changes that were missed by Algorithm \ref{proc1}. The biological roles of those links are not immediately clear, but they point to potentially interesting connectivity patterns that require further scientific validation.

\begin{table}[t!]
\begin{center}
\caption{Significantly different links found by the proposed multiple testing procedure for the KKI-42 dataset. Reported are the significant links identified by the power-enhanced Algorithm \ref{proc2} but missed by Algorithm \ref{proc1} . \label{tab:binary-addition}}
\resizebox{\textwidth}{!}{
\begin{tabular}{ll} \hline
\multicolumn{2}{c}{Differentiating links} \\ \hline
r.rostralanteriorcingulate $\leftrightarrow$ r.superiorparietal & l.caudalmiddlefrontal $\leftrightarrow$ l.fusiform \\
r.fusiform $\leftrightarrow$ r.postcentral & r.lateralorbitofrontal $\leftrightarrow$ l.isthmuscingulate \\
l.corpuscallosum $\leftrightarrow$ l.lingual & r.paracentral $\leftrightarrow$ l.corpuscallosum \\
r.precentral $\leftrightarrow$ l.isthmuscingulate & l.lingual $\leftrightarrow$ l.parsopercularis \\
r.corpuscallosum $\leftrightarrow$ r.inferiortemporal & r.temporalpole $\leftrightarrow$ l.medialorbitofrontal \\
r.lingual $\leftrightarrow$ l.precuneus & r.parsorbitalis $\leftrightarrow$ r.superiortemporal \\
r.corpuscallosum $\leftrightarrow$ r.fusiform & r.fusiform $\leftrightarrow$ r.rostralmiddlefrontal \\
r.lingual $\leftrightarrow$ l.parahippocampal & r.temporalpole $\leftrightarrow$ l.entorhinal \\
r.frontalpole $\leftrightarrow$ l.rostralmiddlefrontal & r.cuneus $\leftrightarrow$ l.pericalcarine \\
l.parahippocampal $\leftrightarrow$ l.pericalcarine & l.isthmuscingulate $\leftrightarrow$ l.frontalpole \\
r.medialorbitofrontal $\leftrightarrow$ l.isthmuscingulate & r.middletemporal $\leftrightarrow$ r.rostralmiddlefrontal \\
r.parsorbitalis $\leftrightarrow$ r.precuneus & l.superiorfrontal $\leftrightarrow$ l.transversetemporal  \\ \hline
\end{tabular}
}
\end{center}
\end{table}

\section{Supplementary Material}
\label{sec:supplement}
The additional lemmas and theorem proofs are available in the online supplementary material.

\bibliographystyle{apa}
\bibliography{ref-nettest}

\begin{thebibliography}{}

\bibitem[\protect\astroncite{Ahn et~al.}{2015}]{Ahn2015}
Ahn, M., Shen, H., Lin, W., and Zhu, H. (2015).
\newblock A sparse reduced rank framework for group analysis of functional
  neuroimaging data.
\newblock {\em Statistica Sinica}, 25:295--312.

\bibitem[\protect\astroncite{Benjamini and Hochberg}{1995}]{BH1995}
Benjamini, Y. and Hochberg, Y. (1995).
\newblock Controlling the false discovery rate: A practical and powerful
  approach to multiple testing.
\newblock {\em Journal of the Royal Statistical Society, Series B.},
  57:289--300.

\bibitem[\protect\astroncite{Bickel et~al.}{2008}]{bickel2008regularized}
Bickel, P.~J., Levina, E., et~al. (2008).
\newblock Regularized estimation of large covariance matrices.
\newblock {\em The Annals of Statistics}, 36:199--227.

\bibitem[\protect\astroncite{Bullmore and Sporns}{2009}]{Bullmore2009}
Bullmore, E. and Sporns, O. (2009).
\newblock {Complex brain networks: graph theoretical analysis of structural and
  functional systems.}
\newblock {\em Nature reviews. Neuroscience}, 10:186--198.

\bibitem[\protect\astroncite{Cai et~al.}{2013}]{cai2013two}
Cai, T.~T., Liu, W., and Xia, Y. (2013).
\newblock Two-sample covariance matrix testing and support recovery in
  high-dimensional and sparse settings.
\newblock {\em Journal of the American Statistical Association}, 108:265--277.

\bibitem[\protect\astroncite{Cai et~al.}{2014}]{cai2014two}
Cai, T.~T., Liu, W., and Xia, Y. (2014).
\newblock Two-sample test of high dimensional means under dependency.
\newblock {\em Journal of the Royal Statistical Society, Series B.},
  76:349--372.

\bibitem[\protect\astroncite{Cai and Zhang}{2016}]{cai2016inference}
Cai, T.~T. and Zhang, A. (2016).
\newblock Inference for high-dimensional differential correlation matrices.
\newblock {\em Journal of multivariate analysis}, 143:107--126.

\bibitem[\protect\astroncite{Chen et~al.}{2015}]{ChenKang2015}
Chen, S., Kang, J., Xing, Y., and Wang, G. (2015).
\newblock A parsimonious statistical method to detect groupwise differentially
  expressed functional connectivity networks.
\newblock {\em Human Brain Mapping}, 36:5196--5206.

\bibitem[\protect\astroncite{Desikan et~al.}{2006}]{desikan2006automated}
Desikan, R.~S., S{\'e}gonne, F., Fischl, B., Quinn, B.~T., Dickerson, B.~C.,
  Blacker, D., Buckner, R.~L., Dale, A.~M., Maguire, R.~P., Hyman, B.~T.,
  et~al. (2006).
\newblock An automated labeling system for subdividing the human cerebral
  cortex on mri scans into gyral based regions of interest.
\newblock {\em Neuroimage}, 31:968--980.

\bibitem[\protect\astroncite{Durante and Dunson}{2018}]{durante2018bayesian}
Durante, D. and Dunson, D.~B. (2018).
\newblock Bayesian inference and testing of group differences in brain
  networks.
\newblock {\em Bayesian Analysis}, 13:29--58.

\bibitem[\protect\astroncite{Fornito et~al.}{2013}]{Fornito2013}
Fornito, A., Zalesky, A., and Breakspear, M. (2013).
\newblock Graph analysis of the human connectome: Promise, progress, and
  pitfalls.
\newblock {\em NeuroImage}, 80:426--444.

\bibitem[\protect\astroncite{Fox and Greicius}{2010}]{Fox2010}
Fox, M.~D. and Greicius, M. (2010).
\newblock Clinical applications of resting state functional connectivity.
\newblock {\em Frontiers in Systems Neuroscience}, 4.

\bibitem[\protect\astroncite{Ginestet et~al.}{2017}]{ginestet2017hypothesis}
Ginestet, C.~E., Li, J., Balachandran, P., Rosenberg, S., Kolaczyk, E.~D.,
  et~al. (2017).
\newblock Hypothesis testing for network data in functional neuroimaging.
\newblock {\em The Annals of Applied Statistics}, 11(2):725--750.

\bibitem[\protect\astroncite{Han et~al.}{2016}]{Han2016}
Han, F., Han, X., Liu, H., and Caffo, B. (2016).
\newblock Sparse median graphs estimation in a high-dimensional semiparametric
  model.
\newblock {\em The Annals of Applied Statistics}, 10:1397--1426.

\bibitem[\protect\astroncite{Kim et~al.}{2014}]{Kim2014}
Kim, J., Wozniak, J.~R., Mueller, B.~A., Shen, X., and Pan, W. (2014).
\newblock Comparison of statistical tests for group differences in brain
  functional networks.
\newblock {\em NeuroImage}, 101:681--694.

\bibitem[\protect\astroncite{Kolaczyk et~al.}{2019}]{kolaczyk2017averages}
Kolaczyk, E., Lin, L., Rosenberg, S., and Walters, J. (2019).
\newblock Averages of unlabeled networks: Geometric characterization and
  asymptotic behavior.
\newblock {\em Annals of Statistics}, to appear.

\bibitem[\protect\astroncite{Landman et~al.}{2011}]{landman2011multi}
Landman, B.~A., Huang, A.~J., Gifford, A., Vikram, D.~S., Lim, I. A.~L.,
  Farrell, J.~A., Bogovic, J.~A., Hua, J., Chen, M., Jarso, S., et~al. (2011).
\newblock Multi-parametric neuroimaging reproducibility: a 3-t resource study.
\newblock {\em Neuroimage}, 54:2854--2866.

\bibitem[\protect\astroncite{Li and Chen}{2012}]{li2012two}
Li, J. and Chen, S.~X. (2012).
\newblock Two sample tests for high-dimensional covariance matrices.
\newblock {\em The Annals of Statistics}, 40:908--940.

\bibitem[\protect\astroncite{Liu}{2013}]{liu2013ggm}
Liu, W. (2013).
\newblock Gaussian graphical model estimation with false discovery rate
  control.
\newblock {\em The Annals of Statistics}, 41:2948--2978.

\bibitem[\protect\astroncite{Luscombe et~al.}{2004}]{Luscombe2004}
Luscombe, N.~M., Madan~Babu, M., Yu, H., Snyder, M., Teichmann, S.~A., and
  Gerstein, M. (2004).
\newblock Genomic analysis of regulatory network dynamics reveals large
  topological changes.
\newblock {\em Nature}, 431:308--312.

\bibitem[\protect\astroncite{Morrison and Hof}{1997}]{MorrisonHof1997}
Morrison, J.~H. and Hof, P.~R. (1997).
\newblock Life and death of neurons in the aging brain.
\newblock {\em Science}, 278:412--419.

\bibitem[\protect\astroncite{Qiu et~al.}{2016}]{QiuHan2016}
Qiu, H., Han, F., Liu, H., and Caffo, B. (2016).
\newblock Joint estimation of multiple graphical models from high dimensional
  time series.
\newblock {\em Journal of Royal Statistical Society, Series B.}, 78:487--504.

\bibitem[\protect\astroncite{Raichle and Gusnard}{2002}]{Raichle2002}
Raichle, M.~E. and Gusnard, D.~A. (2002).
\newblock Appraising the brain's energy budget.
\newblock {\em Proceedings of the National Academy of Sciences},
  99:10237--10239.

\bibitem[\protect\astroncite{Rothman et~al.}{2008}]{rothman2008sparse}
Rothman, A.~J., Bickel, P.~J., Levina, E., and Zhu, J. (2008).
\newblock Sparse permutation invariant covariance estimation.
\newblock {\em Electronic Journal of Statistics}, 2:494--515.

\bibitem[\protect\astroncite{Schott}{2007}]{schott2007some}
Schott, J.~R. (2007).
\newblock Some high-dimensional tests for a one-way {MANOVA}.
\newblock {\em Journal of Multivariate Analysis}, 98:1825--1839.

\bibitem[\protect\astroncite{Schweder and
  Spj{\o}tvoll}{1982}]{schweder1982plots}
Schweder, T. and Spj{\o}tvoll, E. (1982).
\newblock Plots of p-values to evaluate many tests simultaneously.
\newblock {\em Biometrika}, 69:493--502.

\bibitem[\protect\astroncite{Srivastava and
  Yanagihara}{2010}]{srivastava2010testing}
Srivastava, M.~S. and Yanagihara, H. (2010).
\newblock Testing the equality of several covariance matrices with fewer
  observations than the dimension.
\newblock {\em Journal of Multivariate Analysis}, 101:1319--1329.

\bibitem[\protect\astroncite{Storey}{2002}]{storey2002direct}
Storey, J.~D. (2002).
\newblock A direct approach to false discovery rates.
\newblock {\em Journal of the Royal Statistical Society, Series B.},
  64:479--498.

\bibitem[\protect\astroncite{Van~de Geer et~al.}{2014}]{van2014asymptotically}
Van~de Geer, S., B{\"u}hlmann, P., Ritov, Y., Dezeure, R., et~al. (2014).
\newblock On asymptotically optimal confidence regions and tests for
  high-dimensional models.
\newblock {\em The Annals of Statistics}, 42:1166--1202.

\bibitem[\protect\astroncite{Varoquaux and Craddock}{2013}]{Varoquaux2013}
Varoquaux, G. and Craddock, R.~C. (2013).
\newblock Learning and comparing functional connectomes across subjects.
\newblock {\em NeuroImage}, 80:405--415.
\newblock Mapping the Connectome.

\bibitem[\protect\astroncite{Wang et~al.}{2017}]{wang2017common}
Wang, L., Zhang, Z., and Dunson, D. (2017).
\newblock Common and individual structure of multiple networks.
\newblock {\em arXiv preprint arXiv:1707.06360}.

\bibitem[\protect\astroncite{Wang et~al.}{2016}]{WangKang2016}
Wang, Y., Kang, J., Kemmer, P.~B., and Guo, Y. (2016).
\newblock An efficient and reliable statistical method for estimating
  functional connectivity in large scale brain networks using partial
  correlation.
\newblock {\em Frontiers in Neuroscience}, 10:1--17.

\bibitem[\protect\astroncite{Xia et~al.}{2013}]{Xia2013}
Xia, M., Wang, J., and He, Y. (2013).
\newblock Brainnet viewer: A network visualization tool for human brain
  connectomics.
\newblock {\em PLOS ONE}, 8:1--15.

\bibitem[\protect\astroncite{Xia et~al.}{2015}]{xia2015testing}
Xia, Y., Cai, T., and Cai, T.~T. (2015).
\newblock Testing differential networks with applications to the detection of
  gene-gene interactions.
\newblock {\em Biometrika}, 102:247--266.

\bibitem[\protect\astroncite{Xia et~al.}{2019a}]{Xia2018GAP}
Xia, Y., Cai, T.~T., and Sun, W. (2019a).
\newblock {GAP:} a general framework for information pooling in two-sample
  sparse inference.
\newblock {\em Journal of the American Statistical Association}, to appear.

\bibitem[\protect\astroncite{Xia and Li}{2019}]{xia2018two}
Xia, Y. and Li, L. (2019).
\newblock Matrix graph hypothesis testing and application in brain connectivity
  alternation detection.
\newblock {\em Statistica Sinica}, 29:303--328.

\bibitem[\protect\astroncite{Xia et~al.}{2019b}]{xia2019simultaneous}
Xia, Y., Li, L., Lockhart, S.~N., and Jagust, W.~J. (2019b).
\newblock Simultaneous covariance inference for multimodal integrative
  analysis.
\newblock {\em Journal of the American Statistical Association},
  accepted:1--30.

\bibitem[\protect\astroncite{Yuan}{2010}]{yuan2010high}
Yuan, M. (2010).
\newblock High dimensional inverse covariance matrix estimation via linear
  programming.
\newblock {\em Journal of Machine Learning Research}, 11:2261--2286.

\bibitem[\protect\astroncite{Zhang et~al.}{2018}]{Zhang2018}
Zhang, Z., Descoteaux, M., Zhang, J., Girard, G., Chamberland, M., Dunson, D.,
  Srivastava, A., and Zhu, H. (2018).
\newblock Mapping population-based structural connectomes.
\newblock {\em NeuroImage}, 172:130--145.

\bibitem[\protect\astroncite{Zhu and Li}{2018}]{ZhuLi2018}
Zhu, Y. and Li, L. (2018).
\newblock Multiple matrix gaussian graphs estimation.
\newblock {\em Journal of the Royal Statistical Society, Series B.},
  80:927--950.

\end{thebibliography}

\end{document}